\documentclass[prd,english,preprintnumbers,amsmath,amssymb,nofootinbib,superscriptaddress,twocolumn]{revtex4-1}

\pdfoutput=1

\usepackage{mathtools}
\usepackage{amsfonts}
\usepackage{mathrsfs}
\usepackage{bbm}
\usepackage{slashed}

\usepackage{bbold}

\usepackage{graphicx}
\usepackage[dvipsnames]{xcolor}
\usepackage{array}

\usepackage{xspace}
\usepackage{siunitx}
\usepackage{hyperref}

\usepackage{xifthen}
\usepackage{dsfont}
\usepackage{appendix}
\usepackage{enumitem}
\usepackage{booktabs}
\usepackage{units}

\usepackage[latin1]{inputenc}
\usepackage[dvipsnames]{xcolor}
\usepackage{array}

\graphicspath{
	{./}
}

\def\0#1#2{\frac{#1}{#2}}

\def\s0#1#2{\mbox{\small{$ \frac{#1}{#2} $}}}

\def\CC{{\mathcal C}}

\newcommand{\I}{\mathrm{i}}
\newcommand{\be}{\begin{eqnarray}}
\newcommand{\ee}{\end{eqnarray}}

\newcommand{\nn}{\nonumber }

\newcommand{\beq}{\begin{equation}}
\newcommand{\eeq}{\end{equation}}
\newcommand{\bea}{\begin{eqnarray}}
\newcommand{\eea}{\end{eqnarray}}

\newcommand{\psib}{\bar{\psi}}

\def\0#1#2{\frac{#1}{#2}}

\hypersetup{
	colorlinks,
	linkcolor={red!75!black},
	citecolor={blue!75!black},
	urlcolor={blue!75!black},
	pdftitle={Zero-temperature thermodynamics of dense strong-interaction matter},
	pdfauthor={Braun, Schallmo},
	pdfkeywords={Renormalization Group} {Color superconductivity}
	bookmarksopen=true,
	bookmarksopenlevel=2,
	bookmarksnumbered=true
}

\newcommand{\muu}{\mu_\text{u}}
\newcommand{\mud}{\mu_\text{d}}
\newcommand{\mur}{\mu_\text{r}}
\newcommand{\mug}{\mu_\text{g}}
\newcommand{\mub}{\mu_\text{b}}
\newcommand{\mue}{\mu_\text{e}}
\newcommand{\muq}{\mu_\text{Q}}

\usepackage{babel}
\makeatother
\begin{document}

\title{Zero-temperature thermodynamics of dense asymmetric strong-interaction matter}

\author{Jens Braun}
\affiliation{Institut f\"ur Kernphysik, Technische Universit\"at Darmstadt, 
D-64289 Darmstadt, Germany}
\affiliation{ExtreMe Matter Institute EMMI, GSI, Planckstra{\ss}e 1, D-64291 Darmstadt, Germany}
\author{Benedikt Schallmo} 
\affiliation{Institut f\"ur Kernphysik, Technische Universit\"at Darmstadt, 
D-64289 Darmstadt, Germany}

\begin{abstract}
Employing constraints derived from the microscopic theory of the strong interaction, 
we estimate the zero-temperature phase structure 
of dense isospin-asymmetric matter with two quark flavors. We find indications that strong-interaction matter along trajectories relevant 
for astrophysical applications undergoes a first-order phase transition from a color-superconducting phase to an ungapped quark-matter 
phase when the density is increased. Such a phase transition is found to 
be absent in isospin-symmetric matter. Moreover, by taking into account constraints from $\beta$-equilibrium, charge neutrality, and 
color neutrality, we provide an estimate for the speed of sound in neutron-star matter.
Notably, we observe that the speed of sound in neutron-star matter exceeds the asymptotic value associated with the noninteracting quark gas 
and even increases towards lower densities across a wide range, in agreement with recent results for isospin-symmetric matter. Considering results 
from studies based on chiral effective field theory at low densities, our findings suggest the existence of a maximum in the 
speed of sound for~$n/n_0 \lesssim 10$, where~$n_0$ is the nuclear saturation density. 
\end{abstract}
\maketitle

%
\section{Introduction}
\label{sec:intro}
The detection of the gravitational-wave signal of a neutron-star merger~\cite{TheLIGOScientific:2017qsa,Abbott:2018wiz}, the ongoing
missions aiming at direct neutron-star radius measurements~\cite{Watts:2016uzu,NICER,NICER2,miller2021radius,riley2021nicer,Raaijmakers:2021uju}, and 
precise mass measurements of heavy neutron stars~\cite{Demorest10,Antoniadis13,Fonseca2016,2019arXiv190406759C} 
put our understanding of the dynamics of dense matter  to the test, see Ref.~\cite{Huth:2020ozf} for a recent discussion.  
In fact, a quantitative theoretical description of astrophysical objects requires a detailed knowledge of the equation of state (EOS)
of strong-interaction matter over a wide range of densities, up to ten times the nuclear saturation density (or maybe even higher). In addition, for the description of 
neutron-star mergers, information on the temperature dependence of the EOS (up to temperatures of~$\sim 100\,\text{MeV}$) may also become very relevant.
Unfortunately, the theoretical description of strong-interaction matter across a wide density and temperature range is highly 
nontrivial as it requires to bridge the gap between different (effective) degrees of freedom. 

At low densities, chiral effective field theory (EFT) provides a framework 
to describe the properties of nuclear matter in a systematic fashion by means of pions and 
nucleons as low-energy degrees of freedom~\cite{Epelbaum:2008ga}. 
As a consequence, calculations based on chiral EFT interactions yield strong constraints for the low-density part of the EOS~\cite{Hebeler:2013nza,Leonhardt:2019fua}, 
see Ref.~\cite{Hebeler:2020ocj} for a review. For related functional renormalization group (fRG) studies, we refer the reader to 
Refs.~\cite{Berges:1998ha,Kamikado:2012bt,Drews:2014wba,Drews:2014spa,Tripolt:2017zgc,Otto:2019zjy,Otto:2020hoz}.

Going beyond the nucleonic low-density regime, the situation is less clear with respect to the dominant effective degrees of freedom. In fact, an analysis of the 
renormalization group (RG) flow of
gluon-induced four-quark interaction channels in a Fierz-complete setting for two massless quark flavors suggests that 
many interaction channels (including vector channels) become equally strong in a range of densities close to the nucleonic low-density regime~\cite{Leonhardt:2019fua}. 
A quantitative description of the dynamics in this regime therefore requires 
to include at least vector channels, as also discussed in Refs.~\cite{Song:2019qoh,Pisarski:2021aoz,Tripolt:2021jtp}. In fact, the inclusion 
of such channels may even lead to a qualitative change of the EOS in this density regime~\cite{Pisarski:2021aoz}.

Increasing the density even further, the aforementioned Fierz-complete study of gluon-induced four-quark interaction channels suggests the formation of a 
chirally symmetric diquark condensate associated with pairing of the two-flavor color-superconductor (2SC) type, as indicated by a clear dominance of the corresponding 
four-quark channel~\cite{Braun:2019aow}. This channel has also been considered in various early seminal works, ranging from low-energy 
model studies~\cite{Alford:1997zt,Rapp:1997zu,Schafer:1998na,Berges:1998rc} to first-principles studies relying on the fact that the strong coupling becomes small at high densities owing to 
asymptotic freedom~\cite{Son:1998uk,Schafer:1999jg,Pisarski:1999bf,Pisarski:1999tv,Brown:1999aq,Evans:1999at,Hong:1999fh}. 
Given this plethora of studies, it is presently indeed widely accepted that strong-interaction matter at low 
temperatures and sufficiently high densities is a color superconductor with diquarks as effective degrees of freedom, 
see  Refs.~\cite{Bailin:1983bm,Rajagopal:2000wf,Alford:2001dt,Buballa:2003qv,Rischke:2003mt,Shovkovy:2004me,Alford:2007xm,Fukushima:2010bq,%
Fukushima:2011jc,Anglani:2013gfu,Schmitt:2014eka,Baym:2017whm} for reviews. 
Finally, at very high densities and under the assumption that a potentially existing color-superconducting gap 
in the excitation spectrum of the quarks is small compared to the chemical potential, 
constraints on the EOS have been computed in a 
perturbative setting~\cite{Freedman:1976xs,Freedman:1976ub,Baluni:1977ms,Kurkela:2009gj,Fraga:2013qra,Fraga:2016yxs,Gorda:2018gpy,Gorda:2021znl,Gorda:2021kme}. 

The ``detection" of the most relevant effective degrees of freedom in certain density ranges is very 
relevant for astrophysical applications since it opens up the opportunity 
to provide reliable constraints on the EOS of Quantum Chromodynamics (QCD) over a wide density range. Indeed, based on the aforementioned RG analysis of the dominance 
patterns of gluon-induced four-quark interaction channels~\cite{Braun:2019aow}, constraints on the EOS of isospin-symmetric QCD matter with two massless quark flavors 
have been computed~\cite{Leonhardt:2019fua}. Interestingly, the results from this study are not only consistent with 
EOS calculations based on chiral EFT interactions at low densities but also predict the existence of a maximum in the speed of sound at supranuclear densities which exceeds 
the asymptotic high-density value of the speed of sound associated with a noninteracting quark gas. The appearance of this maximum could be traced back to the 
formation of a diquark gap. From an astrophysical standpoint, this is worth mentioning since, at least for neutron-rich matter, the existence of  
a maximum in the speed of sound is presumably required to meet constraints from the analysis of neutron-star masses~\cite{Bedaque:2014sqa,Tews:2018kmu,Greif:2018njt,Annala:2019puf,Huth:2020ozf}.  
Recently, the fRG calculations presented in Ref.~\cite{Leonhardt:2019fua}
have been further developed at intermediate and high densities by taking into account higher-order interactions 
and further resolving the momentum dependence of vertices~\cite{Braun:2021uua}. 
The results from this study are consistent with those from Ref.~\cite{Leonhardt:2019fua}. In particular, first estimates for the speed of 
sound in isospin-symmetric strong-interaction matter are still found to exceed the value 
of the noninteracting quark gas at high densities and even increase as the density is decreased, suggesting the existence of a maximum below $n/n_0\sim 10$, 
where $n_0$ is the nuclear saturation density.

The present work should be viewed as the first extension of  
our series of studies of dense strong-interaction matter~\cite{Braun:2017srn,Braun:2018svj,Braun:2018svj,Braun:2019aow,Leonhardt:2019fua,Braun:2020bhy} to 
finite isospin asymmetry, which is ultimately required to reach out to astrophysical applications. 
In the following, however, we do not aim at a first-principles fRG study of dense isospin-asymmetric QCD matter 
but at a further development of existing low-energy models for QCD with two (massless) quark flavors
at intermediate and high densities by 
taking into account constraints from calculations based on the fundamental quark and gluon degrees of freedom~\cite{Leonhardt:2019fua,Leonhardt:2019fua,Braun:2021uua}, see Sec.~\ref{sec:model}.
Note that, although effects from strange quarks may become relevant at high densities (see Ref.~\cite{Alford:2002kj} for a detailed discussion), 
we shall ignore them in the following for simplicity. Moreover, we shall restrict ourselves to the zero-temperature limit. 
Nevertheless, our present work 
may still provide valuable information on the properties of dense QCD matter and also define a starting point for the inclusion of strange quarks in the future. 
The QCD-constrained low-energy model introduced in Sec.~\ref{sec:model} is used in two ways in Sec.~\ref{sec:daqcd}: In Subsec.~\ref{subsec:pdaqcd}, 
we employ it to estimate the zero-temperature phase diagram of asymmetric QCD 
matter for~$n/n_0 \gtrsim 10$. Based on these results, we then discuss implications for neutron-star matter in 
Subsec.~\ref{subsec:tnstar} by taking into account $\beta$-equilibrium, electric charge neutrality, 
and color neutrality. In particular, we present estimates for the speed of sound.
Our conclusions can be found in Sec.~\ref{sec:conc}.

\section{Model}
\label{sec:model}
In this section, we construct a low-energy model for QCD at intermediate and high densities by exploiting 
results from RG studies of the quark-gluon dynamics in this density regime~\cite{Leonhardt:2019fua,Leonhardt:2019fua,Braun:2021uua}. 
In Subsec.~\ref{sec:gcqcd}, we first discuss general aspects of our model and how it emerges
from quark-gluon interactions in QCD. The effective potential of this model is then derived in Subsec.~\ref{subsec:effact}. 
In Subsec.~\ref{sec:gpqcd}, we present a qualitative discussion of the phase structure and the thermodynamic properties of our model, 
including possible implications for the properties of dense QCD matter. For more quantitative studies, it is required to determine 
the model parameters. In Subsec.~\ref{sec:pcqcd}, we discuss how these parameters can be constrained with information 
from RG flows in QCD.

\subsection{General aspects}
\label{sec:gcqcd}
In QCD, all interaction channels are originally generated by the quark-gluon vertex. 
Whereas the dynamics at high energies can be well described 
in a perturbative setting, the low-energy regime is governed by nonperturbative phenomena, such as 
spontaneous symmetry breaking. Therefore, the description of the low-energy regime by means of suitably 
chosen effective degrees of freedom may be efficient to capture the most relevant dynamics of QCD for a given 
range of the external parameters (e.g., temperature and density). For example, QCD at low densities and temperatures can be well 
described by means of pions and nucleons. At high densities, the dynamics in the long-range limit may then be governed by 
diquarks. The transformation of the effective action of QCD 
under a variation of the resolution and the dynamical emergence of effective degrees of freedom can be 
studied with RG methods~\cite{Gies:2001nw,Gies:2002hq,Pawlowski:2005xe,Gies:2006wv,Braun:2008pi,Floerchinger:2009uf,Braun:2009gm,Braun:2014ata,%
Mitter:2014wpa,Cyrol:2017ewj,Fu:2019hdw,Leonhardt:2019fua,Fukushima:2021ctq,Braun:2021uua}.

A change in the relevant (effective) degrees of freedom suggests to introduce a scale~$\Lambda_{\text{LEM}}$ below which 
a description of QCD by means of a low-energy model~(LEM) constructed from a specific set of effective degrees of freedom 
is suitable. The actual value of~$\Lambda_{\text{LEM}}$ is of minor importance as it is a scheme-dependent quantity. Only 
the existence of this scale matters. Within a given range of external parameters, where the QCD ground state is expected to be governed by spontaneous 
symmetry breaking, the associated symmetry breaking (SB) scale~$k_{\text{SB}}$ may be considered a suitable estimate for~$\Lambda_{\text{LEM}}$. However, 
the scale~$k_{\text{SB}}$ is a priori unknown. Even more, this choice may be impractical as the symmetry breaking scale~$k_{\text{SB}}$ 
may come with an unknown dependence on the external parameters of interest. Therefore, the model scale~$\Lambda_{\text{LEM}}$ is usually 
chosen such that spontaneous symmetry breaking has not yet set in at this scale and, at the same time, the contributions of the 
gauge degrees of freedom to the couplings of the low-energy model become subdominant. Such a range 
of scales may indeed exist. In Refs.~\cite{Braun:2005uj,Braun:2006jd,Braun:2011pp,Leonhardt:2019fua}, for example, it has been discussed 
that four-quark self-interactions generated via two-gluon exchange can become strong enough to contribute as relevant 
operators to the RG flow. Once this is the case, the corresponding four-quark couplings~$\bar{\lambda}_i$ are found 
to increase rapidly, indicating the onset of spontaneous symmetry breaking, see Ref.~\cite{Braun:2011pp} for a detailed discussion. 

An analysis of the RG flow of gluon-induced four-quark interactions above and close to the symmetry breaking scale~$k_{\text{SB}}$ 
can indeed be very useful to gain an insight into the QCD phase structure and also to identify symmetry 
breaking patterns which potentially govern the dynamics at low energies~\cite{Braun:2019aow}. However, a computation 
of, e.g., thermodynamic observables requires to resolve the momentum 
dependence of correlation functions below the symmetry breaking scale. 
In order to gain access to such observables at densities~$n/n_0 \gtrsim 10$, 
we now construct a suitable low-energy model. To this end, we first note that  
a recent RG study of gluon-induced four-quark interactions suggests that 
the diquark channel~$\sim (\psib_b\tau_2\epsilon_{abc}\gamma_5\CC \psib^T_c)(\psi^T_d\CC\gamma_5\tau_2\epsilon_{ade}\psi_e)$ is 
the most dominant channel in this density regime, indicating that 
the formation of a chirally symmetric diquark condensate~$\langle\psib_b\tau_2\epsilon_{abc}\gamma_5\CC \psib^T_c\rangle$ associated with 
pairing of the two-flavor color-superconductor (2SC) type is favored~\cite{Braun:2019aow}.\footnote{Here, we have $\mathcal{C}=\I \gamma_2 \gamma_0$ and the 
Pauli matrix $\tau_2$ is implicitly coupled to the flavor indices of the quarks. In color space, it is 
summed over the totally antisymmetric tensor~$\epsilon_{abc}$.}
We therefore include this four-quark channel in the action~$S_{\text{LEM}}$ of our low-energy model, 
in accordance with early studies of dense strong-interaction matter~\cite{Rapp:1997zu,Alford:1997zt,Berges:1998rc,Pisarski:1999tv,Pisarski:1999bf,Schafer:1999jg}. 
Moreover, to resolve (at least some of) the momentum dependence of this channel, we shall immediately 
rewrite it by introducing auxiliary fields~$\bar{\Delta}_a\sim  (\psib_b\tau_2\epsilon_{abc}\gamma_5\CC \psib^T_c)$:
\be
&& \frac{1}{2}  {\rm i} (\psi^T_b\CC\gamma_5\tau_2\bar{\Delta}_a \epsilon_{abc}\psi_c) 
- \frac{1}{2} {\rm i} (\psib_b\gamma_5\tau_2 \bar{\Delta}^\ast_a \epsilon_{abc}\CC \psib^T_c) \nn\\
&& \qquad\qquad\qquad + \frac{1}{2}\bar{\lambda}_{\text{csc}}^{-1}\bar{\Delta}^\ast_a\bar{\Delta}_a\,.
\label{eq:Yukawa}
\ee
The four-quark coupling~$\bar{\lambda}_{\text{csc}}$ associated with the original four-quark channel then appears as part of the coefficient of the 
curvature term~$\sim\bar{\Delta}^\ast_a\bar{\Delta}_a$.
Note that, in Eq.~\eqref{eq:Yukawa}, we have absorbed the Yukawa coupling~$\bar{h}$ into the (auxiliary) fields~$\bar{\Delta}_a=\bar{h}\Delta_a$. 
This implies that a finite expectation value of the complex-valued field~$\bar{\Delta}_a$ can be identified
with the gap in the fermionic excitation spectrum, which indicates the formation of a color superconductor. 

At first glance, quark self-interaction channels of higher order may be considered parametrically suppressed at high momentum scales. 
For example, eight-quark interactions scale as~$\sim\! {g}^8$, whereas the aforementioned four-quark interactions 
only scale as~$\sim\! {g}^4$. Here,~$g$ denotes the strong coupling. 
In Ref.~\cite{Braun:2021uua}, however, it has been found that, e.g., diquark-like eight-quark interactions
already become relevant above the symmetry breaking scale~$k_{\text{SB}}$, 
i.e., at scales of the order of the model scale~$\Lambda_{\text{LEM}}$. 
Therefore, a corresponding channel should also be included in the (classical) action~$S_{\text{LEM}}$ of our low-energy model. 
As discussed in Ref.~\cite{Braun:2021uua}, this can be conveniently done in the form of a four-diquark channel:
\be
\sim \bar{\lambda}_{\text{eff}} (\bar{\Delta}^\ast_a\bar{\Delta}_a)^2\,,
\label{eq:lambdaeff}
\ee
where~$\bar{\lambda}_{\text{eff}}>0$ is an effective coupling constant. 
In fact, as the curvature term~$\sim\bar{\Delta}^\ast_a\bar{\Delta}_a$, 
this quartic term directly affects the position of the minimum of the effective potential.
Indeed, the associated eight-quark interactions lead to a weaker dependence of the gap on the chemical potential, see 
Ref.~\cite{Braun:2021uua}.

In principle, one may now argue that gluon-induced quark self-interactions of even higher order (e.g., as parameterized in the 
form of diquark interaction channels~$\sim (\bar{\Delta}^\ast_a\bar{\Delta}_a)^{2m}$ with~$m\!>\! 2$) 
could also become relevant at scales of the order of the model scale~$\Lambda_{\text{LEM}}$, 
although such interactions naively scale as~$\sim g^{4m}$ at high momentum scales. 
However, no constraints are presently available for such higher-order interactions. Therefore, 
we do not include them in the (classical) action~$S_{\text{LEM}}$ of our model but only 
take those into account for which constraints are available in Ref.~\cite{Braun:2021uua}, 
see also Subsec.~\ref{sec:pcqcd}.

In order to provide at least potentially useful information for astrophysical applications (e.g., estimates for the density 
dependence of thermodynamic quantities), it 
is required to study matter away from the isospin-symmetric limit. In our model, this is achieved by 
allowing for a difference in the chemical potentials of
the up and down quarks. We shall refer to these flavor chemical potentials as~$\muu$ and~$\mud$, respectively. 
With respect to studies of neutron-star matter, it is also required to 
implement $\beta$-equilibrium which entails the inclusion of a (relativistic) kinetic term for electrons and a corresponding chemical potential~$\mu_{\rm e}$. 
However, we shall neglect quark-electron interactions and interactions among the electrons themselves as 
such interactions are much weaker than those governed by the strong interaction. 

Our low-energy model constructed from the interaction channels~\eqref{eq:Yukawa} and~\eqref{eq:lambdaeff} is not confining. 
While this may not be an issue at high densities, an implementation 
of color neutrality (not to be confused with color confinement) is nevertheless required since, e.g., neutron stars should not carry a net color charge. 
As detailed in Refs.~\cite{Alford:2002kj,Steiner:2002gx,Buballa:2005bv}, we shall implement color neutrality with the aid of two additional 
chemical potentials, $\mu_3$ and~$\mu_8$, coupled to the color charges associated with the generators~$T_3=\lambda_3/2$ and~$T_8=\lambda_8/2$, 
which span the Cartan subalgebra of $\mathfrak{su}(3)$; the $\lambda_a$'s are the Gell-Mann matrices. 

Finally, we note that the fields~$\bar{\Delta}_a$ enter our model as auxiliary fields in the spirit of a 
Hubbard-Stratonovich transformation~\cite{Hubbard:1959ub,Stratonovich}, i.e., they enter the (classical) action~$S_{\text{LEM}}$ 
of our model without a kinetic term. Thus, these fields are not considered to be dynamical 
degrees of freedom at the model scale~$\Lambda_{\text{LEM}}$. Nevertheless, quantum corrections may render them dynamical. 
In the following, we shall ignore such corrections, see also Ref.~\cite{Braun:2021uua} 
for a detailed discussion of this aspect.

\subsection{Effective potential}
\label{subsec:effact}
Guided by the considerations in the preceding subsection, we make the following ansatz for the action of our low-energy model:
{\allowdisplaybreaks
\be
S_{\text{LEM}} &=&\int\text{d}^4x\,\Big\{\psib ({\rm i}\slashed{\partial} - {\rm i}(\hat{\mu}_{\rm (f)}
+ \hat{\mu}_{\rm (c)}))\psi
\nn\\
&&\qquad\qquad
+ \frac{1}{2}\bar{\lambda}_{\text{csc}}^{-1}\bar{\Delta}^\ast_a\bar{\Delta}_a
 + \bar{\lambda}_{\text{eff}} (\bar{\Delta}^\ast_a\bar{\Delta}_a)^2
\nn\\
&&
\qquad\qquad\quad + \frac{1}{2}  {\rm i} (\psi^T_b\CC\gamma_5\tau_2\bar{\Delta}_a \epsilon_{abc}\psi_c) \nn\\
&& 
\qquad\qquad\quad\quad - \frac{1}{2} {\rm i} (\psib_b\gamma_5\tau_2 \bar{\Delta}^\ast_a \epsilon_{abc}\CC \psib^T_c)\nn\\
&& \qquad\qquad\quad\quad\quad +\psib^\text{(e)}({\rm i}\slashed{\partial}- {\rm i}\mue\gamma_0)\psi^\text{(e)}\Big\}\,.
\label{eq:SLEM}
\ee
Here, the} fields~$\psi$ are associated with quarks whereas the fields~$\psi^\text{(e)}$ describe electrons. Explicit 
indices of quark fields refer to their color components. 
For convenience, we have introduced the following auxiliary operators:
\be
\hat{\mu}_{\rm (f)} &=& \text{diag}\big(\muu,\mud\big)_{\rm f}\otimes {\mathbb{1}}_{\text{c}} \otimes \gamma_0 
\label{eq:muf}
\ee
and
\be
\hat{\mu}_{\rm (c)} 
=  {\mathbb{1}}_{\text{f}} \otimes  \text{diag}\big(\mur,\mug,\mub\big)_{\text{c}} \otimes \gamma_0\,,
\ee
where, on the right-hand side, the index~`f' refers to flavor space and `c' to color space. 
The chemical potentials~$\mur$, $\mug$, and~$\mub$ associated with the three color charges (red, green, and blue) are directly related to the 
aforementioned chemical potentials~$\mu_3$ and~$\mu_8$. Indeed, we have 
\be
 \text{diag}\big(\mur,\mug,\mub\big)_{\text{c}} =\left( \mu_3 T_3 + \mu_8 T_8\right) 
\ee
with
\be
\mur &=& \frac{1}{2\sqrt{3}}\mu_8+\frac{1}{2}\mu_3\,,\\
\mug &=& \frac{1}{2\sqrt{3}}\mu_8-\frac{1}{2}\mu_3\,,\\
\mub &=& -\frac{1}{\sqrt{3}}\mu_8\,.
\ee
With respect to the implementation of $\beta$-equilibrium, electric charge neutrality, and color neutrality, we add 
that the chemical potentials~$\muu$, $\mud$, $\mu_{3}$, $\mu_{8}$, and~$\mue$ are not independent 
parameters. We shall come back to this issue in Subsec.~\ref{subsec:tnstar}.

Next, we compute the effective potential~$U$ of our low-energy model in a one-loop approximation 
where we only take into account ``pure" fermion loops.
To this end, we expand the auxiliary fields in Eq.~\eqref{eq:SLEM} 
about a homogeneous background which we choose to point into the 3-direction 
in color space for convenience. By integrating out the quarks and the electrons, we then obtain
\be
 \!\!\!\!\! U&=& \frac{1}{2}\bar{\lambda}_{\text{csc}}^{-1}  |\bar{\Delta}|^2 + 
 \bar{\lambda}_{\text{eff}}  |\bar{\Delta}|^4
 - \frac{\mue^4}{12\pi^2} - \frac{\mu_\text{u,b}^4}{12\pi^2}- \frac{\mu_\text{d,b}^4}{12\pi^2}
   \nn\\
&& \quad  -  8 \bar{l} (|\bar{\Delta}|^2) +  \theta\big(\delta\mu^2 - |\bar{\Delta}|^2\big) \delta \bar{l}(|\bar{\Delta}|^2)\,.
\label{eq:effactmodel}
\ee
Here, $|\bar{\Delta}|^2=\bar{\Delta}_3^{\ast}\bar{\Delta}_3$,\footnote{Physical observables depend only on the (gauge-invariant) 
quantity~$|\bar{\Delta}|^2\equiv \bar{\Delta}_a^{\ast}\bar{\Delta}_a$ (where summation over~$a$ is assumed, $a=1,2,3$). 
Hence, they do not depend on the chosen direction of the background field in color space, see also Ref.~\cite{Rajagopal:2000wf}. 
Moreover, the invariance of the action~$S_{\text{LEM}}$ under color transformations also implies that the effective potential can only be a function of~$|\bar{\Delta}|^2$. 
We have only picked the $3$-direction for convenience which, loosely speaking, implies that the (diquark) condensate is only formed out of ``red and green quarks".} 
where~$\bar{\Delta}_3$ now represents the homogeneous background field.
From a minimization of~$U$ with respect to~$|\bar{\Delta}|^2$, 
we eventually obtain the gap~$\bar{\Delta}_{\text{gap}}$ in the excitation spectrum of the quarks. 

The contributions of the quark loops to the effective potential~$U$ 
depend on~$|\bar{\Delta}|^2$ and 
are parametrized by the functions~$\bar{l}$ and~$\delta \bar{l}$:
\be
\!\!\!\!\! \bar{l}(|\bar{\Delta}|^2) 
&=&\frac{1}{ 4\pi^2} \int_0^\Lambda {\rm d}p\,p^2  \bigg\{ \left( ( p +\bar{\mu})^2 + |\bar{\Delta}|^2 \right)^{\frac{1}{2}}  \nn\\
&&\qquad\qquad\qquad\quad + \left( ( p -\bar{\mu})^2 + |\bar{\Delta}|^2\right)^{\frac{1}{2}}\bigg\} \nn\\
&& \quad- \frac{1}{ 2\pi^2} \int_{\Lambda_{\text{LEM}}}^\Lambda {\rm d}p\,p^2  \bigg\{ \left( p^2 + |\bar{\Delta}|^2\right)^{\frac{1}{2}}  \nn\\
&&\qquad\qquad\qquad\qquad\quad + \frac{\bar{\mu}^2 |\bar{\Delta}|^2}{2( p^2  + |\bar{\Delta}|^2)^{\frac{3}{2}}}
\bigg\} 
\label{eq:defl}
\ee
and
\be
&& \!\!\!\!\!\!\!\!\!\!\! \delta \bar{l}(|\bar{\Delta}|^2)  =  \nn\\
&& \frac{2}{\pi^2} \int_{p_{-}}^{p_{+}}\text{d}p\,p^2  \left(\sqrt{(p- \bar{\mu} )^2+|\bar{\Delta}|^2 }-|\delta\mu|\right)\,,
\label{eq:defdl}
\ee
where 
\be
p_\pm=\bar{\mu} \pm \sqrt{\delta\mu^2 - |\bar{\Delta}|^2}\,.
\ee
We observe that~$\delta\bar{l}$ does not depend on the regularization scale~$\Lambda$ at all. The $\Lambda$-dependence 
of~$\bar{l}$ is removed in the limit~$\Lambda\!\to\! \infty$. Thus, the effective potential~$U$ does also not depend  
on~$\Lambda$ for~$\Lambda\!\to\!\infty$, as it should be (see Ref.~\cite{Braun:2018svj} for a detailed discussion). 
In practice, we ensure the independence of~$\Lambda$ by choosing sufficiently large 
values for~$\Lambda$, i.e.,~$\Lambda \gg  \Lambda_{\text{LEM}} > \bar{\mu}$. 
Note that the actual representations of the functions~$\bar{l}$ and~$\delta\bar{l}$ depend on the details of the regularization. 
In the present work, we employ a three-dimensional sharp cutoff/regulator for convenience. 
However, our results for physical observables do not depend on 
the regulator, provided that the counter terms 
in the third and fourth line of Eq.~\eqref{eq:defl} are chosen carefully. The regulator indeed only affects scheme-dependent quantities, such as  
the values of the model parameters $\bar{\lambda}_{\text{csc}}$ and~$\bar{\lambda}_{\text{eff}}$. 
We add that we simply employ the counter terms derived in Refs.~\cite{Braun:2018svj,Braun:2021uua} since 
these terms are not affected by the isospin asymmetry.

Let us now discuss the dependence of the effective potential~$U$ on the various chemical potentials appearing 
in our model. The contribution~$\sim \mue^4$ in the effective potential~$U$ originates 
from the electrons which only provide a ``charged background" for~$\mue \neq 0$ but do not interact with 
the quarks otherwise.

The various quark chemical potentials enter the effective potential~$U$ only in specific combinations. 
To be precise, we have
\be
\bar{\mu}=\frac{\muu+\mud}{2}+\frac{1}{2\sqrt{3}}\mu_8,\quad
\delta\mu = \frac{\muu-\mud}{2}\,.
\label{eq:bmudmu}
\ee
The quantity~$\bar{\mu}$ may be viewed as the average chemical potential of the two quark flavors. 
The isospin asymmetry can be controlled by the parameter~$\delta\mu$. 
The chemical potential~$\mu_3$ has already been set to zero in the effective potential~$U$. 
Note that this is not an additional approximation but the correct choice
to ensure color neutrality in our studies of neutron-star matter, see App.~\ref{app:cneutral}. 

In our derivation of the effective potential~$U$, we have chosen the background field~$\bar{\Delta}_a$ to point into the 3-direction in color space 
for convenience. Therefore, the ``blue quarks" only appear as ``noninteracting spectators" in 
the effective potential~$U$ via their chemical potentials~$\mu_\text{u,b}$ and~$\mu_\text{d,b}$:
\be
\mu_\text{u,b} =\muu- \frac{1}{\sqrt{3}}\mu_8,\quad 
\mu_\text{d,b} =\mud- \frac{1}{\sqrt{3}}\mu_8\,.
\ee

We close this subsection by summarizing a few general aspects on the computation of thermodynamic 
quantities. First of all, for a given set of chemical potentials, $\vec{\mu}:=(\bar{\mu}, \delta\mu, \mu_3, \mu_8,\mue)$, the 
pressure~$P$ is obtained by evaluating the effective potential~$U$ at its minimum~$|\bar{\Delta}|^2\!=\!|\bar{\Delta}_\text{gap}|^2$:
\be
P=-U \big|_{\text{min},\vec{\mu}} - P_0\,.
\label{eq:press}
\ee
Here, $P_0$ is the {\it vacuum} constant (associated with \mbox{$\vec{\mu}=\vec{0}$}). Since the QCD vacuum   
is governed by spontaneous chiral symmetry breaking, this constant cannot be computed reliably within our diquark model. 
This entails that the pressure is also not accessible in our present study. However, 
towards higher densities (where diquark-like channels are expected to dominate the dynamics at low temperatures~\cite{Braun:2019aow,Leonhardt:2019fua}), 
derivatives of the pressure with respect to the various chemical potentials are accessible. This includes densities and the speed of sound. 
To be specific, we have 
\be
 n_\text{e}=\frac{\partial P}{\partial \mue}
\ee
for the electron density. The densities of the up and down quarks are given by
\be
n_\text{u}=\frac{\partial P}{\partial \muu}\quad\text{and}\quad n_\text{d}=\frac{\partial P}{\partial \mud}\,,
\ee
respectively. Finally, the densities associated with the color charges can be obtained from 
\be
n_3=\frac{\partial P}{\partial \mu_3}\quad\text{and}\quad n_\text{8}=\frac{\partial P}{\partial \mu_8}\,.
\ee
At this point, we would like to add that color neutrality is ensured by choosing the chemical potentials~$\mu_3$ 
and~$\mu_8$ such that the ``color densities"~$n_3$ and~$n_8$ vanish simultaneously. 
As discussed in Subsec.~\ref{subsec:tnstar} and App.~\ref{app:cneutral}, this requires to choose~$\mu_3=0$.

In the Introduction, we have pointed out that the speed of sound~$c_{\rm s}$ is a quantity of particular interest 
in the context of astrophysical applications, see  
Refs.~\cite{Bedaque:2014sqa,Tews:2018kmu,Greif:2018njt,Annala:2019puf,Huth:2020ozf}. This quantity is directly related to the 
first derivative of the pressure with respect to the energy density~$\epsilon$:
\be
c_\text{s} = \left(\frac{\partial P}{\partial \epsilon}\right)^{\frac{1}{2}}\,,
\label{eq:csdef}
\ee
where
\be
\epsilon= -P +\muu n_\text{u} +\mud n_\text{d} +\mue n_\text{e}\,.
\ee
Note that also this quantity does not depend on the vacuum constant~$P_0$ and is therefore accessible within our present work. In Sec.~\ref{sec:daqcd}, 
we shall present estimates for the speed of sound in isospin-symmetric strong-interaction matter,  
charge-neutral strong-interaction matter in $\beta$-equilibrium, and charge- and color-neutral strong-interaction matter in $\beta$-equilibrium.

\subsection{Qualitative discussion of the phase structure and thermodynamics}
\label{sec:gpqcd}
Before we present numerical results for the zero-temperature phase diagram of isospin-asymmetric matter and the speed of sound, we 
discuss the phase structure and the thermodynamics of our model on a qualitative level. For simplicity, we shall restrict our discussion to the case of vanishing 
chemical potentials for the color charges throughout this subsection, i.e., we set~$\mu_3=\mu_8=0$. In any case, this already 
provides us with useful information for the determination of the parameters of our model.

\begin{figure}[t]
\includegraphics[width=\linewidth]{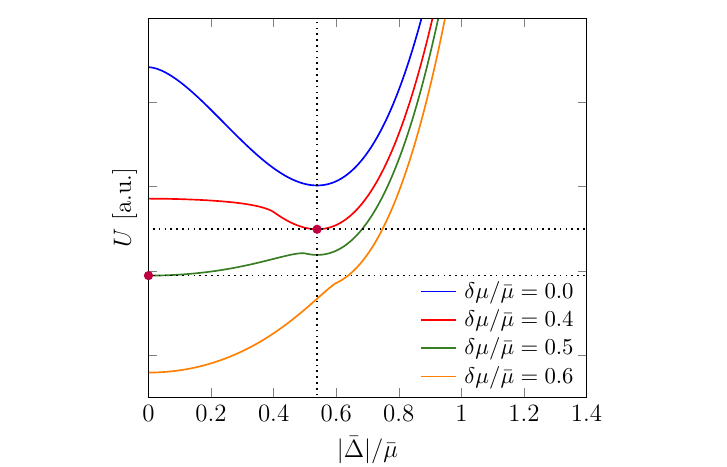}
\caption{\label{fig:pot}Qualitative illustration of the effective 
potential~$U$ for~$\mu_{\rm e}\!=\! 0$ as a function of~$|\bar{\Delta}|/\bar{\mu}$ 
for various values of the relative isospin asymmetry~$\delta\mu/\bar{\mu}$. A discussion of the model parameters 
can be found in Subsec.~\ref{sec:pcqcd}.}
\end{figure}

Let us first consider the effective potential~\eqref{eq:effactmodel} in 
the isospin-symmetric case for~$\bar{\mu}>0$. 
We shall also assume that the parameters~$\bar{\lambda}_{\text{csc}}$ and~$\bar{\lambda}_{\text{eff}}$ have been tuned such that the loop contributions 
parametrized by the functions~$\bar{l}$ and~$\delta\bar{l}$ generate an effective potential~$U$ with a 
nontrivial minimum at~$|\bar{\Delta}|=|\bar{\Delta}_{\text{gap}}(\bar{\mu},\delta\mu\!=\!0)|$, see Fig.~\ref{fig:pot} (blue line) for an illustration. 
Note that the potential is invariant under~$\delta\mu\to -\delta\mu$. Without loss of generality, it therefore suffices 
to consider~$\delta\mu> 0$ in this subsection.

It is now important to observe that the quantum correction~$\delta\bar{l}$ does not contribute to the effective potential for~$\delta\mu=0$. Even more,
we deduce from Eq.~\eqref{eq:effactmodel} that an increase 
of the isospin-asymmetry parameter~$\delta\mu$ does not affect the dependence of the effective potential on~$|\bar{\Delta}|^2$ 
in the domain~$|\bar{\Delta}|^2 > \delta\mu^2$. In particular, this implies that the position of the 
minimum remains unchanged for~$\delta\mu^2 < |\bar{\Delta}_{\text{gap}}(\bar{\mu},\delta\mu\!=\!0)|^2$.
In other words, the original minimum of the effective potential is not shifted and remains to be a 
minimum for all~$\delta\mu^2 < |\bar{\Delta}_{\text{gap}}(\bar{\mu},\delta\mu\!=\!0)|^2$. However, by evaluating the effective potential, 
we also observe that the~$\delta\mu$-dependent quark-loop contribution~$\delta\bar{l}$ in Eq.~\eqref{eq:effactmodel} generates a second minimum at~$|\bar{\Delta}|=0$ for sufficiently 
large values of~$\delta\mu$, see Fig.~\ref{fig:pot} (green line) for an illustration. 
These observations suggest that a critical value~$\delta\mu_{\text{cr}}\!>\! 0$ of the isospin-asymmetry parameter~$\delta\mu$ may exist 
such that the minimum at~$|\bar{\Delta}|=|\bar{\Delta}_{\text{gap}}(\bar{\mu},\delta\mu\!=\!0)|$ is no longer the global minimum of the effective potential for~$\delta\mu^2 > \delta\mu_{\text{cr}}^2$, 
but may only be a local minimum, where~$\delta\mu_{\text{cr}}^2 < |\bar{\Delta}_{\text{gap}}(\bar{\mu},\delta\mu\!=\!0)|^2$. At~$\delta\mu = \delta\mu_{\text{cr}}$, the system 
should then undergo a first-order phase transition. In any case, 
these considerations indicate the existence of a finite range of values of~$\delta\mu$ for which  
the gap~$\bar{\Delta}_{\text{gap}}(\bar{\mu},\delta\mu)$ is identical to the gap~$\bar{\Delta}_{\text{gap}}(\bar{\mu},\delta\mu\!=\!0)$ in the isospin-symmetric limit, 
see also Ref.~\cite{Bedaque:1999nu} for a corresponding discussion.

We add that not all quark degrees of freedom contribute to the formation of a color-superconducting ground state. For example, choosing   
the homogeneous background to point into the $3$-direction in color space as done in this work, the ``blue 
quarks" eventually only appear as ``noninteracting spectators" in the effective potential, see Eq.~\eqref{eq:effactmodel}. 
Note that, within our present approximations, a corresponding 
contribution to the effective potential is expected for any choice of the homogeneous background. 

Let us now discuss the existence of a critical asymmetry~$\delta\mu_{\text{cr}}$ for the chemical potentials 
from a more general standpoint. The existence of such a critical value is indeed a very 
well-known feature of two-component Fermi gases with a superconducting ground state.
For example, in the case of an electron gas, a magnetic field can be used to generate an asymmetry between 
``spin-up and spin-down electrons" because of the Zeeman coupling of the electron spin to the magnetic field. In other words, the magnetic 
field can be used to polarize the system. When the magnetic field becomes sufficiently strong, superfluidity 
is found to disappear and the system undergoes a first-order phase transition to an ungapped partially polarized phase, 
as shown in the seminal works of 
Chandrasekhar~\cite{Chandrasekhar} and Clogston~\cite{Clogston}. This first-order phase transition is associated with a critical polarization. 
The search for a corresponding critical polarization in ultracold atomic two-component Femi gases in the unitary limit has also attracted a lot of 
attention in the past 10-15 years, see Refs.~\cite{Chevy2010,Zwerger2012,GUBBELS2013255} for reviews. 

Following the works of Chandrasekhar and Clogston, the existence of a critical polarization can be understood from a 
comparison of the pressure of the unpolarized superfluid (gapped) phase with the pressure of the ungapped partially polarized phase. 
Searching for the ground state is then equivalent to searching for the phase with highest pressure (i.e., lowest Gibbs energy).  

In principle, a similar line of arguments can also be applied to QCD.
Loosely speaking, the formation of diquarks from quarks in the color-superconducting 
phase yields an energy gain per quark which is of the order of the gap $\bar{\Delta}_{\text{gap}}(\bar{\mu},\delta\mu\!=\!0)$ in the isospin-symmetric limit. 
Note that the diquarks in our model are composed of one up and one down quark. 
Color-superconductivity then disappears, if the energy~$\delta\mu$ gained by adding, e.g., an up quark to the system (rather than a down quark) 
exceeds a critical value~$\delta\mu_{\text{cr}}$. The scale for~$\delta\mu_{\text{cr}}$ is set by the gap $\bar{\Delta}_{\text{gap}}(\bar{\mu},\delta\mu\!=\!0)$.
An estimate for~$\delta\mu_{\text{cr}}$ can be obtained by comparing the pressure of the isospin-asymmetric 
quark gas in the noninteracting limit (Stefan-Boltzmann limit),
\be
P_{\text{SB}}(\bar{\mu},\delta\mu)=\frac{(\bar{\mu}+\delta\mu)^4}{4\pi^2} + \frac{(\bar{\mu}-\delta\mu)^4}{4\pi^2}\,,
\label{eq:PSBasym}
\ee
with the pressure of the gapped isospin-symmetric phase evaluated at leading 
order in the gap~\cite{Rajagopal:2000wf,Rajagopal:2000ff,Shovkovy:2002kv,Braun:2018svj,Braun:2021uua}:\footnote{This expression 
can also be derived from Eq.~\eqref{eq:effactmodel} in the limit of vanishing four-diquark coupling. Moreover, 
at leading order in~$\bar{\Delta}_{\text{gap}}(\bar{\mu},0)/\bar{\mu}$, the general form of Eq.~\eqref{eq:Pwclimit} 
may already be deduced from purely dimensional arguments in QCD since the gap sets the scale at high densities, 
see also Ref.~\cite{Braun:2022jme} for a discussion.}
\be
P =  P_{\text{SB}}(\bar{\mu},0) \left( 1 + 2 \left(\frac{\bar{\Delta}_{\text{gap}}(\bar{\mu},0)}{\bar{\mu}}\right)^2 + \dots\right) \,.
\label{eq:Pwclimit}
\ee
From this comparison, we obtain
\be
\delta\mu_{\text{cr}}(\bar{\mu}) = \frac{1}{\sqrt{3}} \bar{\Delta}_{\text{gap}}(\bar{\mu},0) + \dots
\label{eq:critasym}
\ee
for the critical isospin asymmetry.\footnote{Here, terms of higher order in the gap have been dropped.} We observe 
that~$\delta\mu_{\text{cr}}$ inherits the~$\bar{\mu}$-dependence of the gap in the isospin-symmetric limit.\footnote{Taking into account 
that effectively only two of the three color degrees of freedom of the quarks are gapped (see our discussion below), we 
obtain~$\delta\mu_{\text{cr}}(\bar{\mu}) = (1/\sqrt{2}) \bar{\Delta}_{\text{gap}}(\bar{\mu},0) + \dots$. This can be seen 
by {\it not} taking into account the contributions from the ungapped quarks into the analysis leading to Eq.~\eqref{eq:critasym}.} 
Moreover, since the pressure is not continuously differentiable at~$\delta\mu_{\text{cr}}$ for a given fixed~$\bar{\mu}$, 
we expect the system to undergo a first-order phase transition at this point.
A similar analysis for QCD with $2+1$ quark flavors
can be found in Refs.~\cite{Rajagopal:2000ff,Alford:2017ale}.

One may now be tempted to conclude that all physical observables remain equal to their values in the isospin-symmetric limit 
for~$\delta\mu^2 < \delta\mu_{\text{cr}}^2$. In particular, we may expect the pressure and the gap to remain constant within this range of values of~$\delta\mu$.
However, this is not the case, since, loosely speaking, only some quark degrees of freedom contribute to the formation of the color-superconducting ground 
state in our model, see our discussion above. The remaining other quark degrees do not couple to the (diquark) condensate.  
To be more specific, the formation of a finite expectation value of the 
form~$\sim\langle \psib_b\tau_2\epsilon_{abc}\gamma_5\CC \psib^T_c\rangle$ (with fixed~$a$)  
is associated with the symmetry breaking pattern~$\text{SU}(3)\to \text{SU}(2)$ in color space. 
For example, choosing~$a=3$ for convenience (as done in all explicit calculations 
in the present work), a gap~$\bar{\Delta}_{\text{gap}}$ is only generated in the subspace of the red and green quarks. 
This subspace is invariant under~$\text{SU}(2)$ color transformations. The blue quarks 
remain ungapped. 
The observed independence of the gap on~$\delta\mu$ below some critical value~$\delta\mu_{\text{cr}}$ then results 
from the position of the poles of the propagators of the red and green quarks in 
the complex $p_0$ plane relative to the real $p_0$ axis 
for given values of the chemical potentials, the spatial loop momentum, and the 
homogeneous diquark background~$\bar{\Delta}_a$.\footnote{Here,~$p_0$ refers to the zeroth component of the four-momentum 
appearing in the quark propagators.}
We emphasize that the appearance of a gap in the excitation spectrum of the quarks as well as the 
invariance of the gap under a shift of~$\delta\mu$ (for~$\delta\mu^2 < \delta\mu_{\text{cr}}^2$)
are physical statements which may also be present in QCD although 
the local gauge invariance under~$\text{SU}(3)$ color transformations  
cannot be broken~\cite{Elitzur:1975im}. In this respect, we also add that the effective potential~\eqref{eq:effactmodel} in general depends only 
on the gauge-invariant quantity~$\bar{\Delta}_a^{\ast}\bar{\Delta}_a$ (summation over~$a$ is assumed). 
These more general considerations also indicate that thermodynamic quantities (e.g., pressure and densities) 
should still exhibit a dependence on~$\delta\mu$ for~$\delta\mu^2< \delta\mu_{\text{cr}}^2$ which 
originates from the ungapped quarks. In our numerical results for the speed of sound and the densities 
we indeed observe a mild dependence on the isospin-asymmetry parameter in the gapped phase, see Subsec.~\ref{subsec:tnstar}.

We close this subsection by noting that the situation encountered in the aforementioned ultracold unitary Fermi gases is slightly different. 
In these gases, all physical observables indeed remain equal to their values in the gapped phase up to 
the critical asymmetry since both fermion degrees of freedom contribute to the formation of the superconducting ground state.
Note that the  phase transition is found to be of first order in these gases, even if fluctuation 
effects are taken into account~\cite{Boettcher:2014tfa,Boettcher:2014xna,Roscher:2015xha,Frank_2018}.  

\subsection{Parameter constraints from QCD}
\label{sec:pcqcd}
Let us begin our discussion of the determination of the model 
parameters~$\bar{\lambda}_{\text{csc}}$ and~$\bar{\lambda}_{\text{eff}}$ by considering the isospin-symmetric limit. 
In Ref.~\cite{Braun:2021uua}, an analysis of RG flows starting from the QCD action indicated that the four-quark coupling~$\bar{\lambda}_{\text{csc}}$  
in the isospin-symmetric limit depends only very mildly on the chemical potential, at least at RG scales~$k$  
(sufficiently) greater than the symmetry breaking scale~$k_{\text{SB}}$. Since we choose~$\Lambda_{\text{LEM}}>k_{\text{SB}}$, 
we shall assume that~$\bar{\lambda}_{\text{csc}}$ does not depend on~$\bar{\mu}$. 

The situation is different for the effective four-diquark coupling~$\bar{\lambda}_{\text{eff}}$. 
Indeed, the aforementioned study of RG flows in QCD suggests that this coupling depends on the 
chemical potential~$\bar{\mu}$, even above the symmetry breaking scale. 
In principle, one may now exploit the RG flows in Ref.~\cite{Braun:2021uua} to extract the values of the couplings~$\bar{\lambda}_{\text{csc}}$  
and~$\bar{\lambda}_{\text{eff}}$ at a suitably chosen model scale~$\Lambda_{\text{LEM}}$, which could then be used 
as input parameters for the evaluation of the effective potential~\eqref{eq:effactmodel} of our model. 
However, the values of couplings are 
scheme-dependent quantities which  in general renders such a matching of coupling values complicated. In our present work, 
we therefore tune the model parameters~$\bar{\lambda}_{\text{csc}}$ and~$\bar{\lambda}_{\text{eff}}$ for a given value of the 
chemical potential such that we recover the value of the gap found 
in Ref.~\cite{Braun:2021uua}. Note that the gap in the excitation spectrum of the quarks is a physical observable. 

\begin{figure}[t]
\includegraphics[width=\linewidth]{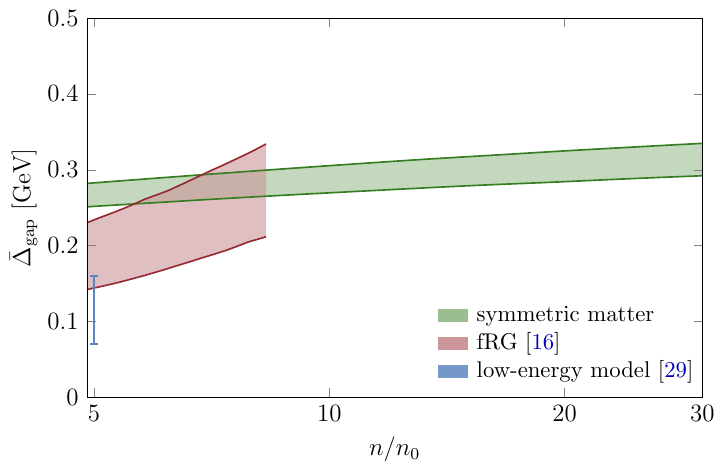}
\caption{\label{fig:gap}Gap in the excitation spectrum of the quarks in isospin-symmetric matter 
as a function of the baryon density~$n$ in units of the nuclear saturation density~$n_0$. The green-shaded band represents the gap 
used in the present work as input parameter (based on results from RG flows in  QCD~\cite{Braun:2021uua}). 
The red-shaded band depicts results from an fRG study~\cite{Leonhardt:2019fua}. As an example 
for conventional low-energy models studies, we also show the result for the gap from Ref.~\cite{Alford:1997zt}.}
\end{figure}

To be specific, we choose~$\Lambda_{\text{LEM}}=1\,\text{GeV}$ and restrict ourselves 
to regimes with~$\bar{\mu}<\Lambda_{\text{LEM}}$. For the four-quark coupling, we 
use~$\bar{\lambda}_{\text{csc}}^{-1}\approx 0.197\,\text{GeV}^2$ (for all values of~$\bar{\mu}$).
As a function of~$\bar{\mu}$, the remaining model parameter~$\bar{\lambda}_{\text{eff}}$ is then tuned such that 
the gap in our present work agrees with the one found in the study of QCD RG flows in Ref.~\cite{Braun:2021uua}. 
Since the gap in this study comes with an uncertainty band originating from a variation of the RG scheme 
and an uncertainty in the strong coupling, we employ two parameter sets for~$\bar{\lambda}_{\text{eff}}$ 
which are associated with the lower and upper end 
of the uncertainty band of the gap, respectively. The corresponding values 
for~$\bar{\lambda}_{\text{eff}}$ can be found in Ref.~\cite{Braun:2021uua}. 
In Fig.~\ref{fig:gap} (green band), we show the resulting gap  
as a function of the baryon density~$n$ in units of the nuclear saturation density~$n_0$. 

At this point, we would like to comment on the behavior of the gap as a function of the density. In Ref.~\cite{Braun:2021uua}, 
a comparatively weak dependence of the gap on the chemical potential~($\sim$ density) has been found, see 
green-shaded band in Fig.~\ref{fig:gap}. 
The rapid flattening of the gap with increasing density can be traced back to the 
emergence of the four-diquark coupling already above the symmetry breaking scale.   
For~$n/n_0\gtrsim 7$, the results for the gap from Ref.~\cite{Braun:2021uua} (which we have used to fix our model parameters) 
are consistent with those obtained based on a Fierz-complete study of gluon-induced four-quark interaction channels~\cite{Leonhardt:2019fua}.
This is in accordance with the fact that the latter study predicts the diquark interaction channel to be most dominant. 
However, for lower densities, roughly~$n/n_0\lesssim 10$, interaction channels other than the diquark channel become 
very relevant~\cite{Braun:2019aow}. In particular, the chiral scalar-pseudoscalar channel starts to play a dominant role and 
leads to a stronger decrease of the gap than observed in Ref.~\cite{Braun:2021uua} for decreasing density, see red-shaded band in Fig.~\ref{fig:gap}.
From a comparison of the red-shaded and green-shaded band, 
we also cautiously conclude that the regime~$n/n_0 \lesssim 7$ is no longer accessible in our present work. 

In Fig.~\ref{fig:gap}, we observe that the gap from Ref.~\cite{Leonhardt:2019fua} is greater than the one reported in 
previous low-energy model studies, see, e.g., Ref.~\cite{Alford:1997zt}. This may be traced back 
to the quark-gluon dynamics underlying the study in Ref.~\cite{Leonhardt:2019fua}. Indeed, the quark-gluon dynamics 
may yield a rapid increase of the gap at 
lower densities,~$\bar{\Delta}_{\text{gap}}\sim \exp(-c/(g^4\bar{\mu}^2))$ (with $c>0$ 
being a constant)~\cite{Braun:2021uua}. 
Nevertheless, the results for the gap from Refs.~\cite{Leonhardt:2019fua} and~\cite{Alford:1997zt} are still in 
accordance at lower densities.

Up to this point, we have basically only discussed how the parameters of our model can be fixed in 
the isospin-symmetric limit. However, we are aiming 
at studies of isospin-asymmetric strong-interaction matter. Since we use the gap to fix our 
model parameters, we can exploit the fact that the gap in our model is independent of the isospin-asymmetry 
parameter~$\delta\mu$ for~$\delta\mu^2 < \delta\mu_{\text{cr}}^2$, see our discussion in Subsec.~\ref{sec:gpqcd} and Fig.~\ref{fig:pot}. 
For~$\delta\mu^2 > \delta\mu_{\text{cr}}^2$, the ground state is then described by a noninteracting isospin-asymmetric 
quark gas. Therefore, it is consistent to use the same 
model parameters for the isospin-symmetric and the isospin-asymmetric case. Of course, the uncertainty in the isospin-symmetric gap   
leads to an uncertainty in our estimates for the phase structure and thermodynamics of isospin-asymmetric matter, see Sec.~\ref{sec:daqcd}. 
This provides an insight into the sensitivity of our results for isospin-asymmetric strong-interaction matter on the uncertainty in the gap. 

\section{Dense asymmetric QCD matter}
\label{sec:daqcd}
\subsection{Phase diagram of dense QCD matter}
\label{subsec:pdaqcd}
Since the computation of thermodynamic quantities with functional approaches also requires the computation 
of the order parameter, we begin our discussion of dense isospin-asymmetric strong-interaction matter by considering the zero-temperature 
phase diagram in the plane spanned by the total baryon density~$n$ (in units of the nuclear saturation density~$n_0$) and 
the down-quark fraction $n_{\rm d}/(n_{\rm u} + n_{\rm d})$. Throughout this subsection, we shall always consider the case of vanishing color 
chemical potentials~$\mu_3$ and~$\mu_8$. Constraints from color neutrality are then discussed in Subsec.~\ref{subsec:tnstar}.

\begin{figure}[t]
\includegraphics[width=\columnwidth]{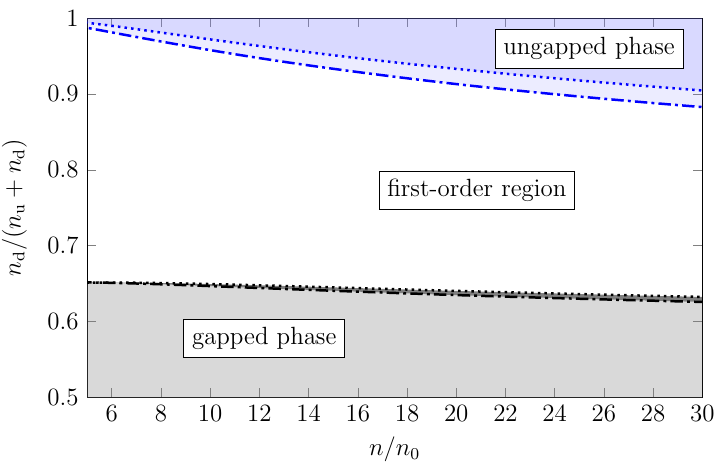}
\caption{\label{fig:pd}Phase diagram of isospin-asymmetric strong-interaction matter in the plane spanned 
by the total baryon density~$n$ (in units of the nuclear saturation density~$n_0$) and 
the down-quark fraction~$n_{\rm d}/(n_{\rm u} + n_{\rm d})$. The uncertainty in the phase boundaries as depicted by 
different line styles results from the uncertainty in the gap (see Fig.~\ref{fig:gap}), where the 
dotted and dashed-dotted lines are associated with the upper and lower end of the uncertainty band of the gap, respectively. 
}
\end{figure}

In Fig.~\ref{fig:pd}, we show the phase diagram spanned by~$n/n_0$ and the down-quark fraction $n_{\rm d}/(n_{\rm u} + n_{\rm d})$ 
for~$5\leq n/n_0 \leq 30$, as obtained from an evaluation of the effective potential~\eqref{eq:effactmodel}. 
In this phase diagram we can identify three different regions, a region associated with a gapped phase (gray-shaded area), 
a region associated with an ungapped phase (blue-shaded area), and a first-order region between these two phases. 
Whereas the gapped phase describes a color-superconductor and is governed by spontaneous symmetry breaking, the 
ungapped phase can in principle be treated in a perturbative 
setting~\cite{Freedman:1976xs,Freedman:1976ub,Baluni:1977ms,Kurkela:2009gj,Fraga:2013qra,Fraga:2016yxs,Gorda:2018gpy}. 
The first-order region is unstable. To be more specific, we observe that the system undergoes a (strong) first-order 
phase transition from the gapped phase to the ungapped phase when we increase the down-quark fraction 
from~$n_{\rm d}/(n_{\rm u} + n_{\rm d})= 1/2$ (symmetric matter) to~$n_{\rm d}/(n_{\rm u} + n_{\rm d})= 1$ (pure down-quark matter) 
for given fixed total density~$n/n_0$.

From Fig.~\ref{fig:pd} we deduce that the phase boundaries decrease with increasing total density~$n/n_0$. This 
can already be understood from a more general standpoint based on our analytic estimate of the critical 
asymmetry~$|\delta\mu_{\text{cr}}|= |\bar{\Delta}_{\text{gap}}(\bar{\mu},\delta\mu\!=\! 0)|/c_{\delta} + \dots$ with~$c_{\delta}>0$, 
see Eq.~\eqref{eq:critasym}.\footnote{For our qualitative analysis in the following, only the sign of the constant~$c_{\delta}$ matters. However, a 
quantitative estimate for this quantity can already be extracted from our analytic calculations above, see Eq.~\eqref{eq:critasym}, where we have found~$c_{\delta}=\sqrt{3}$.}
 Indeed, since we have
\be
\frac{n_{\rm d}}{n_{\rm u} + n_{\rm d}}= \frac{1}{2} - \frac{3}{2}\frac{\delta\mu}{\bar{\mu}} + \dots
\ee
for the down-quark fraction in the ungapped phase for~$\delta\mu/\bar{\mu} \ll 1$, the 
critical down-quark fraction can be expanded as follows
\be
\left(\frac{n_{\rm d}}{n_{\rm u} + n_{\rm d}}\right)_{\text{cr}} = \frac{1}{2} - \frac{3}{2c_{\delta}\bar{\mu}}\bar{\Delta}_{\text{gap}}(\bar{\mu},0) + \dots\,.
\ee
Using now 
\be
n=\frac{1}{3}\left( n_{\rm u}+n_{\rm d}\right) = \frac{2\bar{\mu}^3}{3\pi^2} \left( 1 + 3 \left(\frac{\delta\mu}{\bar{\mu}}\right)^2\right)
\ee
for the total baryon density in the ungapped phase and that the gap scales 
as~$\bar{\Delta}_{\text{gap}} \sim \exp(-c^{\prime}/\bar{\mu}^2)$ with~$c^{\prime}>0$,\footnote{This is the conventional  
scaling behavior expected for the color-superconducting gap in (relativistic) models~\cite{Alford:1997zt,Rapp:1997zu,Schafer:1998na,Berges:1998rc}.    
In QCD, the factor~$c^{\prime}$ depends on the strong coupling~$g$, see our discussion in Subsec.~\ref{sec:pcqcd}. With respect to the gap used as input 
in our present study, see Fig.~\ref{fig:gap}, a fit of the constant~$c^{\prime}$ within the density range relevant for the present work 
yields~$c^{\prime}\approx 0.039\,\text{GeV}^{2}$ (upper end of the green-shaded band) and~$c^{\prime}\approx 0.036\,\text{GeV}^{2}$ (lower end of the green-shaded band).}
we find that the critical down-quark fraction scales as 
\be
\left(\frac{n_{\rm d}}{n_{\rm u} + n_{\rm d}}\right)_{\text{cr}} - \frac{1}{2} \sim \frac{1}{n^{\frac{1}{3}}}\exp\left(- \frac{c^{\prime\prime}}{ n^{\frac{2}{3}}}\right)\,
\label{eq:scale1st}
\ee 
in the high-density limit, where~$c^{\prime\prime}>0$ is a constant.\footnote{Here, we have used that the pressure 
is a continuous function of the chemical potentials, even in the presence of a first-order transition. 
Strictly speaking, the relation~\eqref{eq:scale1st} describes 
the scaling of the upper end of the first-order region shown in Fig.~\ref{fig:pd}. Since the line describing the lower 
end of the first-order region is bounded from above by the upper end, it is reasonable to expect that the lower end of 
this region exhibits a similar scaling behavior, at least at high densities.} Thus, the first-order region approaches the isospin-symmetric line 
(i.e., $n_{\rm d}/(n_{\rm u} + n_{\rm d})=1/2$) for~$n\to\infty$ and the extent of the gapped 
phase shrinks to zero in this limit. However, this does not imply that the gap along the isospin-symmetric line also tends to zero.

\begin{figure}[t]
\includegraphics[width=\columnwidth]{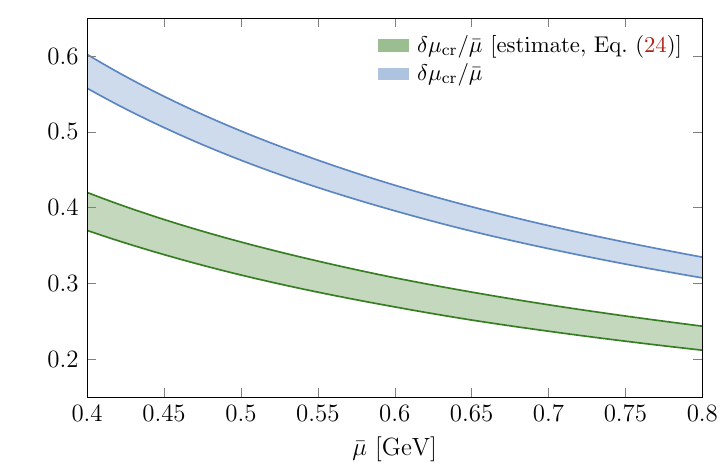}
\caption{\label{fig:mucr}Critical isospin asymmetry~$\delta\mu_{\text{cr}}$ (in units of~$\bar{\mu}$) as a function of 
the average chemical potential~$\bar{\mu}$. We observe that the results for this quantity obtained 
from a minimization of the effective potential~\eqref{eq:effactmodel} agree qualitatively with those from our 
analytic estimate~\eqref{eq:critasym}.}
\end{figure}

We add that the relation~\eqref{eq:scale1st} for the scaling behavior of the phase boundaries is very general. 
In fact, we did not make use of specific properties of our model but only 
relied on two general assumptions: (i) the existence of a color-superconducting gap in QCD and its scaling behavior,\footnote{The details 
of the density dependence of the critical down-quark fraction in 
Eq.~\eqref{eq:scale1st} are determined by the dependence of the gap on the chemical potential 
which may differ from the one assumed in the derivation of Eq.~\eqref{eq:scale1st}, see 
Refs.~\cite{Evans:1998ek,Evans:1998nf,Schafer:1998na,Rischke:2003mt,Alford:2007xm,Braun:2021uua} for discussions of the 
scaling behavior of the gap in QCD. However, the 
statement that the phase boundaries approach the isospin-symmetric line is more general as it only relies on the assumption that~$\bar{\Delta}_{\text{gap}}(\bar{\mu},0)/\bar{\mu}\to 0$
for~$\bar{\mu}\to \infty$ (corresponding to $n\to \infty$). The approach to the isospin-symmetric limit may indeed be even 
slower than described by Eq.~\eqref{eq:scale1st}. For example, the phase boundaries may scale logarithmically as dictated by the scale 
dependence of the strong coupling~$g$.}
and (ii) the general considerations which already led us to the scaling behavior of the critical isospin 
asymmetry~$\delta\mu_{\text{cr}}$. For completeness, we show the 
analytic estimate~\eqref{eq:critasym} for~$\delta\mu_{\text{cr}}$ together with the result extracted 
from the effective potential~\eqref{eq:effactmodel} in Fig.~\ref{fig:mucr}. To evaluate 
the analytic estimate for~$\delta\mu_{\text{cr}}$, we have employed the gap extracted from Eq.~\eqref{eq:effactmodel}. 

Specific trajectories in the phase diagram spanned by the total baryon density~$n$ and 
the down-quark fraction $n_{\rm d}/(n_{\rm u} + n_{\rm d})$ are of particular interest. For example, 
we may want to consider neutron-like matter which corresponds to the 
horizontal line~$n_{\rm d}/(n_{\rm u} + n_{\rm d})=2/3$ in this phase diagram. From Fig.~\ref{fig:pd} we deduce that 
such a neutron-like matter trajectory is located in the unstable first-order region, at least within the considered density range. 
Towards lower densities, this trajectory may then enter the gapped phase before it finally ends up in a phase associated with 
spontaneous chiral symmetry breaking in the low-density regime. As discussed above, however, this low-density regime is not accessible 
with our present model. More importantly with respect to astrophysical applications, we shall also see in the next subsection 
that realistic neutron-star matter trajectories may indeed traverse the gapped phase.
\subsection{Towards neutron-star matter}
\label{subsec:tnstar}
In our present work, we do not include effects from strange quarks. Nevertheless, 
we would like to identify trajectories in the phase diagram spanned by the total baryon density~$n$ and 
the down-quark fraction $n_{\rm d}/(n_{\rm u} + n_{\rm d})$ which are relevant for astrophysical applications, 
at least in a world with only two quark flavors. This requires to 
take into account constraints from $\beta$-equilibrium, electric charge neutrality, and color neutrality (not to be confused with color confinement). 
At least potentially, this may already provide an insight into the thermodynamic properties of neutron-star matter. 
\begin{figure}[t]
\includegraphics[width=\columnwidth]{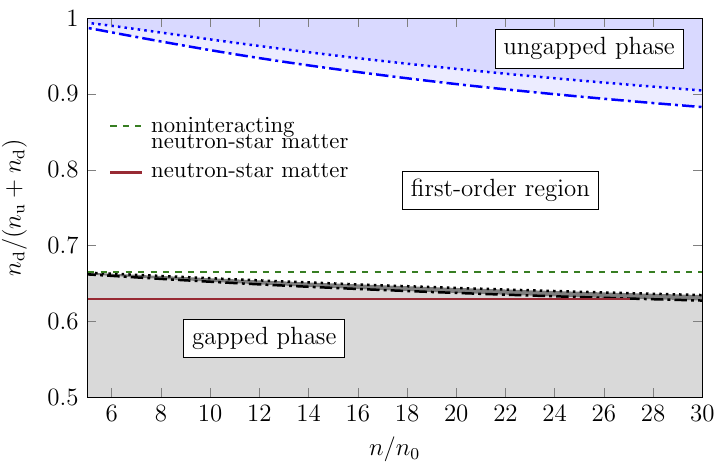}
\caption{\label{fig:pd2}Phase diagram of isospin-asymmetric strong-interaction matter spanned 
by the total baryon density~$n$ (in units of the nuclear saturation density~$n_0$) and 
the down-quark fraction~$n_{\rm d}/(n_{\rm u} + n_{\rm d})$. In contrast to Fig.~\ref{fig:pd}, the phase boundaries now describe
the points where color-neutral matter undergoes a first-order phase transition. 
As it is the case in Fig.~\ref{fig:pd}, the uncertainty in these boundaries (as depicted by different line styles) result from 
the uncertainty in the gap (see Fig.~\ref{fig:gap}). Dotted (dashed-dotted) lines are associated with the upper (lower) 
end of the uncertainty band of the gap. The red (solid) line depicts the neutron-star matter trajectory as obtained from our 
model (taking into account electric charge neutrality, $\beta$-equilibrium, and color neutrality), whereas  
the green dashed line depicts the neutron-star matter trajectory in case of a system of noninteracting quarks.
}
\end{figure}

Neutron-star matter is in $\beta$-equilibrium, see, e.g., Refs.~\cite{Rajagopal:2000ff,Alford:2002kj,Steiner:2002gx}. For our present work, 
this implies that there is no preferred direction for weak-interaction processes of the following type:
\be
u\longleftrightarrow d + e^{+} + \nu_{\rm e}\,.
\ee
In chemical equilibrium the chemical potentials of the particles involved in such processes are then related:
\be
\muu=\mud+\muq\,.
\label{eq:mubeq}
\ee
Here,~$\muq$ is the chemical potential associated with the electric charge~$Q$ which is directly related to the electron chemical potential~$\mue$:
\be
\muq = -\mue\,.
\ee
In Eq.~\eqref{eq:mubeq}, we have set the chemical potential of the neutrinos~$\nu_{\rm e}$ to zero since we assume that they leave the neutron star, see, e.g., 
Refs.~\cite{Alford:2002kj,Steiner:2002gx,Buballa:2003qv,Ebert:2005wr,Andersen:2007qv,Alford:2017ale} for detailed discussions of this aspect.

For a study of physical observables in $\beta$-equilibrium, it is now convenient to rewrite 
the chemical potentials of the two quark flavors by introducing an auxiliary chemical potential~$\mu$ for the quarks:
\be
\muu = \mu +\frac{2}{3}\muq\quad\text{and}\quad
\mud = \mu -\frac{1}{3}\muq\,.
\ee
Here, the prefactors of~$\muq$ are determined by the electric charge of the up and down quarks, respectively. 
With these relations at hand, we can also rewrite the effective potential~$U$. To be specific, 
the chemical potentials~$\bar{\mu}$ and~$\delta\mu$ in Eq.~\eqref{eq:effactmodel} can be expressed
as a sum of the chemical potentials~$\mu$,~$\muq$, and the color chemical potential~$\mu_8$: 
\be
\bar{\mu}=\mu+\frac{1}{6}\muq+\frac{1}{2\sqrt{3}}\mu_8
\ee
and
\be
\delta\mu =\frac{\muu-\mud}{2}= \frac{1}{2}\muq\,.
\ee
The pressure~\eqref{eq:press} then becomes a function of the chemical potentials~$\mu$,~$\muq$, and~$\mu_8$. 
Recall that we have already set~$\mu_3=0$, see below. 

Electric charge neutrality can now be implemented by
requiring that the charge density~$n_{\rm Q}$ vanishes:
\be
n_{\rm Q}=  \frac{\partial P}{\partial \muq}=\frac{2}{3}n_\text{u}-\frac{1}{3}n_\text{d}-n_\text{e}\stackrel{!}{=}0\,.
\label{eq:nQconstr}
\ee
In our present model,~$n_{\rm e}$ is simply the density of a free relativistic electron gas,~$n_{\rm e}=\mu_{\rm e}^3/(3\pi^2)$. 
In the phase diagram spanned by the total baryon density~$n$ and the down-quark fraction $n_{\rm d}/(n_{\rm u} + n_{\rm d})$, 
a trajectory respecting the constraints from $\beta$-equilibrium and electric charge neutrality can be found by minimizing 
the effective potential~\eqref{eq:effactmodel} with respect to~$|\bar{\Delta}|^2$ for a range of values of~$\mu$,~$\muq$, and~$\mu_8$ and then singling out the 
value of the electron chemical potential~$\mue=-\muq$ which fulfills the constraint~\eqref{eq:nQconstr} for given values of~$\mu$ 
and~$\mu_8$. For example, considering~$\mu_8=0$ (i.e., ignoring color neutrality), the electron chemical potential~$\mue$ 
and the chemical potentials of the up and down quarks are then determined by the 
chemical potential~$\mu$. The total density~$n$ as well as the densities~$n_{\rm u}$ and~$n_{\rm d}$ are finally obtained by taking 
derivatives of the pressure with respect to the corresponding chemical potentials on this trajectory.

In addition to charge neutrality and $\beta$-equilibrium, color neutrality should be imposed in a realistic description of 
neutron-star matter~\cite{Alford:2002kj,Steiner:2002gx}. To this end, we initially included the color chemical potentials~$\mu_3$ 
and~$\mu_8$ in the action~\eqref{eq:SLEM} of our model, which allow us to identify a color-neutral trajectory 
in the phase diagram spanned by the total baryon density~$n$ and the down-quark fraction $n_{\rm d}/(n_{\rm u} + n_{\rm d})$ 
with the aid of the following two constraints:
\be
n_8=\frac{\partial P}{\partial\mu_8}=\frac{1}{2\sqrt{3}}\Big(n_\text{r}+n_\text{g}-2n_\text{b}\Big)\stackrel{!}{=}0
\label{eq:n8constr}
\ee
and
\be
n_3=\frac{\partial P}{\partial\mu_3}=\frac{1}{2}\Big(n_\text{r}-n_\text{g}\Big)\stackrel{!}{=}0\,.
\ee
Here, the densities~$n_{\rm r}$, $n_{\rm g}$, and~$n_{\rm b}$ refer to the densities of the 
red, blue, and green quarks, respectively. We observe that~$n_{\rm r}=n_{\rm g}=n_{\rm b}$ is indeed 
ensured by fulfilling these two constraints.

As indicated in Subsec.~\ref{subsec:effact}, 
we have already set~\mbox{$\mu_3=0$} in the effective potential~$U$, see Eq.~\eqref{eq:effactmodel}. 
This is not an additional approximation but follows from the constraints on~$n_3$ and~$n_8$ and entails 
that~$n_3=0$, as detailed in App.~\ref{app:cneutral}. 
Thus, we are left with the chemical potential~$\mu_8$ which can be determined with the aid of constraint~\eqref{eq:n8constr}. 

\begin{figure}[t]
\includegraphics[width=\columnwidth]{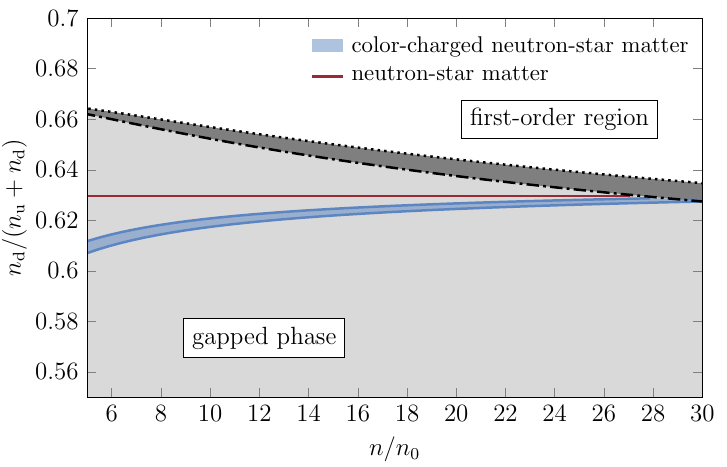}
\caption{\label{fig:nocolor}Phase diagram as shown in Fig.~\ref{fig:pd2} but with a reduced range 
of the axis associated with the down-quark fraction to show the difference between   
the (color-neutral) neutron-star matter trajectory (red line) and the color-charged neutron-star matter trajectory (blue band).}
\end{figure}

\begin{figure}[t]
\includegraphics[width=\columnwidth]{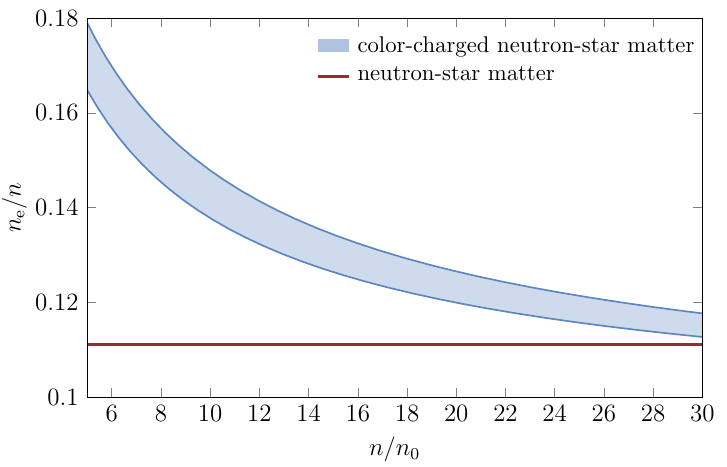}
\caption{\label{fig:efrac}Electron fraction~$n_{\rm e}/n$ 
as a function of the total baryon density~$n$ (in units of the nuclear saturation density~$n_0$) along the (color-neutral) neutron-star matter trajectory and 
the color-charged neutron-star matter trajectory. 
The lower (upper) end of the uncertainty band is associated with the lower (upper) end of the uncertainty band of the gap.}
\end{figure}

In the plane spanned by the total density~$n$ and the down-quark fraction $n_{\rm d}/(n_{\rm u} + n_{\rm d})$, we can now determine a trajectory 
which respects the constraints from color neutrality, charge neutrality and $\beta$-equilibrium. In the following, we shall refer to this trajectory as 
``neutron-star matter trajectory". The computation of this trajectory requires to simultaneously solve Eq.~\eqref{eq:nQconstr}, charge neutrality constraint, 
and Eq.~\eqref{eq:n8constr}, remaining color neutrality constraint. Notably, we obtain the following exact solution for the neutron-star matter trajectory:
\be
\frac{n_\text{d}}{n_\text{u}+n_\text{d}}=\frac{17}{27}\,.
\ee
This down-quark fraction can be translated into a ratio of protons and neutrons which are the most relevant 
effective degrees of freedom at low densities. We find~$n_{\rm p}/n_{\rm n} =1/8$, where~$n_{\rm n}$ 
is the neutron density and~$n_{\rm p}$ is the proton density. For the proton fraction~$n_{\rm p}/(n_{\rm p} + n_{\rm n})$, 
we thus have~$n_{\rm p}/(n_{\rm p} + n_{\rm n})=1/9$, which is identical to the electron fraction:~$n_{\rm e}/n= n_{\rm p}/(n_{\rm p} + n_{\rm n})=1/9$. 
Here,~$n=(n_{\rm u} + n_{\rm d})/3$ is the total baryon density.\footnote{In contrast to most of our results, our results  
for the neutron-star matter trajectory and the corresponding ``proton fraction" do not come with an uncertainty band. 
This can be traced back to the fact that terms suffering from the 
uncertainty in the gap cancel out in the process of solving Eqs.~\eqref{eq:nQconstr} and~\eqref{eq:n8constr} simultaneously. If we ignore 
the constraint~\eqref{eq:n8constr} from color neutrality and simply set~$\mu_3=\mu_8=0$, this is no longer the case, see Figs.~\ref{fig:nocolor} and~\ref{fig:efrac}.}

In Fig.~\ref{fig:pd2}, we show the neutron-star matter trajectory (red solid line) in the phase diagram spanned 
by the baryon density~$n$ and the down-quark fraction~$n_{\rm d}/(n_{\rm u} + n_{\rm d})$. 
For comparison, we also show the neutron-star matter trajectory 
as obtained from the consideration of a noninteracting quark gas (green dashed line in Fig.~\ref{fig:pd2}). Note that the position of the boundaries of the gapped and 
ungapped phase differ slightly from those shown in Fig.~\ref{fig:pd}. This difference results from the fact that the phase boundaries in Fig.~\ref{fig:pd2} 
correspond to points where color-neutral matter undergoes a first-order transition. 

Interestingly, we observe in Fig.~\ref{fig:pd2} that the neutron-star matter trajectory 
traverses the gapped phase but eventually ``hits" the phase boundary at~$n/n_0 \gtrsim 27$. Close to the phase boundary, 
the gap turns out to be still sizeable within our present approximations,~$\bar{\Delta}_{\text{gap}}/\bar{\mu} \sim 0.3$. 
At the transition point, the system then undergoes a (strong) first-order phase transition from 
a gapped phase to an ungapped phase which may be fully accessible in a conventional perturbative setting. 
Note that the existence of such a phase transition may be a model-independent feature. In fact, 
our general considerations in Subsec.~\ref{subsec:pdaqcd} suggest that the phase boundary of the gapped phase 
approaches the line associated with isospin-symmetric matter for~$n\to\infty$, 
see Eq.~\eqref{eq:scale1st}. Thus, one may at least naively expect that neutron-star matter undergoes 
such a phase transition, provided that the corresponding trajectory is located within the gapped phase 
at some point. We emphasize that 
this phase transition should not be confused with a transition from a phase governed by spontaneous chiral 
symmetry breaking to a chirally 
symmetric color-superconducting (gapped) phase which is expected to occur at much lower densities. 

At this point, we also would like to add that there is no regime/``subphase" in Figs.~\ref{fig:pd} and~\ref{fig:pd2} (as well as in Fig.~\ref{fig:nocolor} to be discussed below),
where, loosely speaking, modes with a gap in the isospin-symmetric limit are rendered gapless in the ground state by the presence of a finite isospin imbalance.\footnote{In 
condensed-matter physics, it is referred to such a regime as Sarma phase~\cite{Sarma}, see, e.g., Refs.~\cite{Chevy2010,Zwerger2012,GUBBELS2013255} for reviews in the 
context of ultracold Fermi gases.} 
In particular, we observe that this type of gapless modes is not present along the neutron-star matter trajectory. 
The potential existence of such gapless modes in QCD matter  
has been discussed in detail in Refs.~\cite{Shovkovy:2003uu,Huang:2003xd}, see, e.g., Refs.~\cite{Buballa:2003qv,Shovkovy:2004me} for reviews.

It is instructive to investigate how color neutrality affects our results.\footnote{As discussed above, we expect neutron-star matter to be ``color neutral", 
see Refs.~\cite{Amore:2001uf,Alford:2002kj,Steiner:2002gx,Buballa:2005bv}. Results for ``color charged" neutron-star matter are shown {\it only} to illustrate which of the quantities 
considered in the present work are strongly affected by the implementation of the color-neutrality constraint, at least within our present model setup which allows a study of 
both cases at comparatively low numerical costs. For numerically more intense studies of properties of strong-interaction matter, such a comparison may be potentially useful since  
an implementation of additional constraints (here, the color-neutrality constraint) may then be very costly from a numerical standpoint.}
This can be done by computing the trajectory along which 
charge neutrality and $\beta$-equilibrium are taken into account but the color-neutrality constraints are dropped (i.e.,~$\mu_3\!=\!\mu_8\!=\!0$). 
In the following, we shall refer to this trajectory as ``color-charged neutron-star matter trajectory". In Fig.~\ref{fig:nocolor}, this trajectory is shown 
together with the (color-neutral) neutron-star matter trajectory in the plane spanned by the total density and the 
down-quark fraction. We find that both trajectories lie close to each other and effectively converge when the 
density is increased. As a consequence, the electron fraction~$n_{\rm e}/n$ along the color-charged neutron-star matter trajectory also approaches 
the one along the (color-neutral) neutron-star matter trajectory, see Fig.~\ref{fig:efrac}. However, towards lower densities, we observe that 
the electron fractions associated with the two trajectories start to deviate clearly. Recent constraints from nuclear physics and 
observations disfavor large electron fractions at low densities in neutron stars and also 
suggest that the electron fraction does not increase towards lower densities~\cite{Huth:2020ozf}.
With respect to astrophysical applications, we may therefore cautiously conclude from our results that 
the implementation of color neutrality becomes increasingly relevant when the low-density regime is approached. In any case, the electron fractions 
along both trajectories are still consistent with the aforementioned constraints from nuclear physics and observations. 

\begin{figure}[t]
\includegraphics[width=\columnwidth]{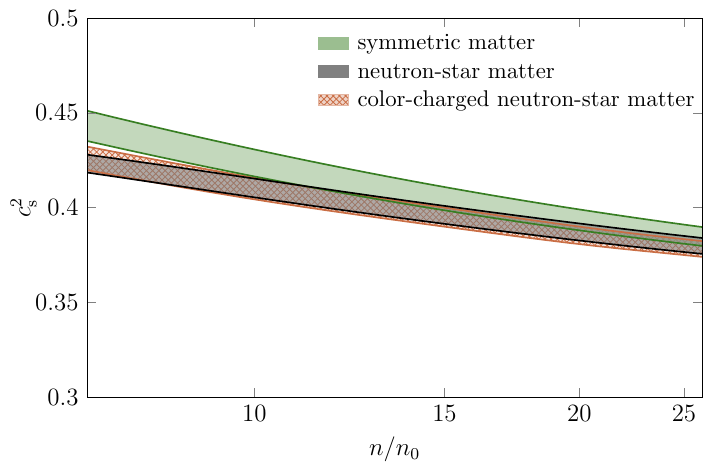}
\caption{\label{fig:css} Speed of sound squared (in units of the speed of light squared) in (isospin-)symmetric matter (green-shaded band), 
color-charged neutron-star matter (red diamond pattern), and (color-neutral) neutron-star matter (black-shaded band) as 
a function of the total baryon density~$n$ (from $n/n_0=7$ to $n/n_0=26$, where $n_0$ is the nuclear saturation density). 
 The lower (upper) ends of the uncertainty bands are associated with the lower (upper) end of the uncertainty band of the gap.}
\end{figure}

Let us finally consider the speed of sound as another quantity of great interest with respect to astrophysical applications. 
To compute the speed of sound along different trajectories, we employ Eq.~\eqref{eq:csdef}. In Fig.~\ref{fig:css}, 
we present our results for the speed of sound squared as a function of the total baryon density along the (color-neutral) neutron-star matter trajectory, 
the trajectory associated with color-charged neutron-star matter in $\beta$-equilibrium, and isospin-symmetric matter (not color-neutral). 
Note that we only show the speed of sound within the gapped phase in Fig.~\ref{fig:css}. Above~$n/n_0\gtrsim 27$, the 
neutron-star matter trajectory as well as the color-charged neutron-star matter trajectory cross the phase boundary, see Figs.~\ref{fig:pd} and~\ref{fig:pd2}. 
Within the gapped phase, we observe that the three trajectories lead to very similar results for the speed of sound. 
In particular, in all three cases, the speed of sound exceeds the value of the noninteracting quark gas and even increases towards lower densities 
within the considered density range.  
It is worth mentioning that, on a qualitative level, this behavior of the speed of sound 
as a function of the density can also be derived analytically from Eq.~\eqref{eq:Pwclimit}.\footnote{For a detailed discussion of corrections 
to Eq.~\eqref{eq:Pwclimit} and how they may affect the speed of sound, we refer the reader to Ref.~\cite{Braun:2022jme}.}
For example, assuming~$\bar{\Delta}_{\text{gap}} \sim \exp(-c^{\prime}/\bar{\mu}^2)$ and~$\bar{\mu}\sim n^{\frac{1}{3}}$ as 
in Subsec.~\ref{subsec:pdaqcd}, we obtain the following estimate for the deviation of the speed of sound squared 
from its value in the noninteracting limit at high densities:
\be
c_{\rm s}^{2}-\frac{1}{3} \sim \frac{1}{n^{\frac{2}{3}}}\exp\left(-\frac{2c^{\prime\prime}}{n^{2/3}}\right)\,.
\label{eq:cssana}
\ee
Here, we have restricted ourselves to the isospin-symmetric case and also dropped corrections 
of higher order in~$1/n^{1/3}$ and~$(\bar{\Delta}_{\text{gap}}(\bar{\mu},0)/\bar{\mu})$.\footnote{For simplicity, we have 
assumed that $\bar{\mu}=(\pi^2 n/2)^{1/3}$ (relation for the noninteracting quark gas). This yields the following relation between the constants~$c^{\prime}$ 
and~$c^{\prime\prime}$: $c^{\prime\prime}=c^{\prime}/(\pi^2/2)^{2/3}$. Of course, within the gapped phase, this is only an approximation. A more quantitative 
estimate may be obtained by employing an ansatz of the following form: $\bar{\mu} = c_{\bar{\mu}}n^{c_{n}}$, where~$c_{\bar{\mu}}$ and 
$c_{n}$ are positive constants. Thus, the actual functional dependence of the density may differ from our present estimate. Still, it appears reasonable to expect that~$c_{n} \approx 1/3$ 
for large~$\bar{\mu}$ where the gap becomes small compared to~$\bar{\mu}$. Note also that the contribution of the ungapped ``blue" quarks to the density scales as~$n^{1/3}$.}
In accordance with our numerical results, we deduce from this relation that the speed of sound indeed approaches its value in the noninteracting limit from above 
at high densities. The latter statement appears to be insensitive to the details of our assumption for the $\bar{\mu}$-dependence of the gap. 
Indeed, provided that the pressure can be written in the form of Eq.~\eqref{eq:Pwclimit} for small~$\bar{\Delta}_{\text{gap}}(\bar{\mu},0)/\bar{\mu}$, 
it is sufficient that~$\bar{\Delta}_{\text{gap}}(\bar{\mu},0)/\bar{\mu}\to 0$ for~$\bar{\mu}\to \infty$ and~$\bar{\Delta}_{\text{gap}}(\bar{\mu},0)$ 
increases monotonically as a function of~$\bar{\mu}$. 

In Fig.~\ref{fig:css}, we also observe that the speed of sound along the neutron-star matter trajectory essentially agrees with the one along the  
color-charged neutron-star matter trajectory for the considered densities within the uncertainty bands. Recall that the electron fractions associated with these 
two trajectories start to deviate clearly towards lower densities, see Fig.~\ref{fig:efrac}. We also deduce from 
Fig.~\ref{fig:css} that a finite isospin asymmetry has the tendency to slightly lower the speed of sound. However, the effect of a finite isospin 
asymmetry diminishes for increasing density. This does not come unexpected.
In fact, our general discussion in Subsec.~\ref{sec:gcqcd} already suggests that the isospin dependence may at least be weak 
for some observables within the color-superconducting phase. For the considered densities, 
this observation may provide a justification to restrict more advanced computations of 
thermodynamic quantities of astrophysical relevance to the isospin-symmetric limit, at least in a first step. 
\begin{figure}[t]
\includegraphics[width=\columnwidth]{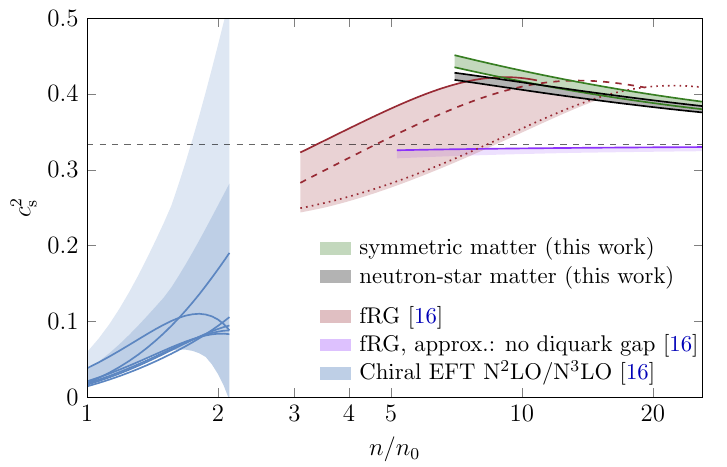}
\caption{\label{fig:css2} Speed of sound squared (in units of the speed of light squared) in (isospin-)symmetric 
matter (green-shaded band, this work and Ref.~\cite{Braun:2021uua}) and in 
neutron-star matter (black-shaded band, this work) as a function of~$n/n_0$, 
together with results for the speed of sound in (isospin-)symmetric matter 
as obtained from calculations based on chiral EFT (blue-shaded bands)~\cite{Leonhardt:2019fua}, 
an fRG study taking into account the formation of a diquark gap (red-shaded band)~\cite{Leonhardt:2019fua}, 
and an fRG study based on an approximation without taking into account a diquark gap~\cite{Leonhardt:2019fua}. 
The black dashed line represents the speed of sound squared in a noninteracting quark gas.}
\end{figure}

In Fig.~\ref{fig:css2}, we finally compare our results for the speed of sound with those from 
a computation based on an fRG analysis of a Fierz-complete set of (gluon-induced) 
four-quark interactions~\cite{Leonhardt:2019fua}. We observe that 
our present results are consistent with those from Ref.~\cite{Leonhardt:2019fua} for \mbox{$n/n_0\gtrsim 7$}.
Taking also into account recent results from studies 
based on chiral EFT interactions at low densities~\cite{Hebeler:2013nza,Leonhardt:2019fua}, the 
scaling behavior of the speed of sound at high densities suggests the existence of a maximum in the speed of sound 
for~$n/n_0 \lesssim 10$, see Fig.~\ref{fig:css2}. Within the existing uncertainties, this is in accordance with the aforementioned Fierz-complete RG study starting 
from the QCD action~\cite{Leonhardt:2019fua}. Note that the existence of an increase of the speed of sound above the value associated with the 
noninteracting quark gas has also been observed and discussed in (low-energy) models where strong-interaction matter is 
studied coming from low densities (see, e.g., Refs.~\cite{McLerran:2018hbz,Pisarski:2021aoz,Contrera:2022tqh,Ivanytskyi:2022oxv,Kojo:2020ztt})
rather than from high densities as done in the present work. Interestingly, such an increase of the speed of sound has also been observed in 
a recent lattice study of two-color QCD~\cite{Iida:2022hyy}.

With respect to astrophysical applications, we add that constraints from 
neutron-star masses strongly suggest the existence of such a maximum in 
neutron-rich matter~\cite{Bedaque:2014sqa,Tews:2018kmu,Greif:2018njt,Annala:2019puf,Huth:2020ozf}.
However, a computation of the position of this maximum is beyond the scope of the present work 
as it will most likely require the inclusion of additional effective degrees of freedom, 
such as pions, vector mesons, and nucleons, which are expected to become 
relevant in this density regime~\cite{McLerran:2018hbz,Braun:2018bik,Braun:2019aow,Song:2019qoh,Pisarski:2021aoz}.

Of course, another quantity of great relevance for astrophysical applications is the pressure as a function of the density. Up to a constant (vacuum constant)
this quantity can be obtained by an integration of the speed of sound squared with respect to the energy density, see Eq.~\eqref{eq:csdef}. In (low-energy) model 
studies, different ways of determining this constant have been discussed, see, e.g., Refs.~\cite{McLerran:2018hbz,Contrera:2022tqh,Ivanytskyi:2022oxv,Kojo:2020ztt}. 
In the following, we shall not determine this constant but only note that a combination of our present work with Refs.~\cite{Leonhardt:2019fua,Braun:2021uua} may be used 
in the future to compute estimates of the pressure and the speed of sound of isospin-imbalanced strong-interaction matter by studying RG flows starting 
from the QCD action at high momentum scales. Results for the pressure of symmetric matter from such an RG study can be found in Ref.~\cite{Leonhardt:2019fua} 
where also a comparison of the pressure with results from model studies is provided. Our present study of the speed of sound (see Figs.~\ref{fig:css} and~\ref{fig:css2}) suggests that 
the results for the pressure presented in Ref.~\cite{Leonhardt:2019fua} may depend only mildly on the isospin asymmetry at high densities, 
at least for imbalances relevant for astrophysical applications. However, a detailed analysis of this aspect is deferred to future work.

\section{Conclusions}
\label{sec:conc}
In order to study the zero-temperature phase diagram of strong-interaction matter with two massless quark flavors at supranuclear densities, 
we have constructed a model based on constraints 
from studies of RG flows in QCD~\cite{Braun:2019aow,Leonhardt:2019fua,Braun:2021uua}. In this model, 
the dynamics is found to be governed by the formation of 
a color-superconducting gap at sufficiently small down-quark fractions.
By increasing the down-quark fraction for a fixed total density, we have found that the system 
undergoes a (strong) first-order phase transition to an ungapped quark-matter phase. 
Our results suggest that, if this phase transition occurs in two-flavor QCD, the associated phase boundary in the 
phase diagram spanned by the total baryon density and the down-quark fraction approaches the 
isospin-symmetric line when the density is increased.

The results for the zero-temperature phase diagram of our model 
are in accordance with more general considerations which do not rely on specific details of our model, 
see our discussion in Subsec.~\ref{subsec:pdaqcd}. These considerations already suggest that 
dense neutron-star matter may undergo a first-order phase transition from a 
color-superconducting phase to an ungapped quark matter phase. 
However, in terms of the density, the position of this transition to the ungapped phase may be well 
beyond densities relevant for astrophysical applications. 
Note that our analysis also indicates that isospin-symmetric strong-interaction 
matter does not undergo such a phase transition at high densities. 

By taking into account constraints from $\beta$-equilibrium, charge neutrality, and color neutrality in our model,
we have determined a neutron-star matter trajectory in the phase diagram spanned by the total baryon density and the down-quark fraction.
This trajectory is found to be located in the gapped phase for~$n/n_0\lesssim 27$ but ``hits" the  
boundary of this phase towards higher densities, as suggested by our general considerations. 
Along this trajectory (but below the phase transition), 
we have computed two quantities of particular interest in the context of astrophysical applications, 
the electron fraction and the speed of sound. Our results for both quantities are in accordance with constraints from 
nuclear physics and observations~\cite{Huth:2020ozf}. In particular, we have found that the speed of sound along the neutron-star matter trajectory 
exceeds the asymptotic value associated with the noninteracting quark gas
and even increases towards lower densities across a wide range.  
Taking into account results from studies based on chiral EFT interactions at low densities~\cite{Hebeler:2013nza,Leonhardt:2019fua}, 
the observed behavior of the speed of sound at~$n/n_0\gtrsim 10$ 
suggests the existence of a maximum in this quantity at intermediate densities, as also discussed 
for isospin-symmetric matter in Ref.~\cite{Leonhardt:2019fua}. 
Note that constraints from neutron-star masses also strongly support the existence of a maximum in the speed of sound in
neutron-rich matter~\cite{Bedaque:2014sqa,Tews:2018kmu,Greif:2018njt,Annala:2019puf,Huth:2020ozf}. 

With respect to astrophysical applications, it should be added that strange quarks may become  
relevant in the density regime considered in this work. In fact, taking strange quarks into account, 
the ground state associated with pairing of the two-flavor color-superconductor type (as considered in this work) may 
no longer be favored~\cite{Alford:2002kj}. Still, for example, our general considerations regarding the density dependence of the 
speed of sound do not rely on the specific type of the gap but only on the dependence of the pressure on the gap
and may therefore also hold in the presence of strange quarks, at least in case of color-flavor locking at high densities.\footnote{We refer to App.~\ref{app:2plus1} for a comment on this aspect.}
In any case, the construction of a $2\!+\!1$-flavor model based on, e.g., 
constraints from RG flows in QCD is in principle also possible but appears to be much more challenging. For example, 
the number of possible channels in a Fierz-complete analysis of gluon-induced four-quark interactions 
already grows drastically because of the explicit breaking of the flavor symmetry~\cite{Braun:2020mhk}. 
Leaving this ambitious endeavor aside, our present work may still help to gain a better understanding 
of the properties of dense strong-interaction matter in general. 

{\it Acknowledgments.--} 
The authors would like to thank M.~Buballa, J.~E.~Drut, A.~Gei\ss el, T.~Gorda, K.~Hebeler, J.~M.~Pawlowski, D.~Rischke, and A.~Schwenk for useful discussions and comments on the manuscript. 
As members of the fQCD collaboration~\cite{fQCD}, the authors also would like to thank the other members of this collaboration for discussions and 
providing data for cross-checks. 
J.B. acknowledges support by the DFG under grant BR~4005/4-1 and BR~4005/6-1 (Heisenberg program).
This work is supported by the Deutsche Forschungsgemeinschaft (DFG, German 1355 Research Foundation) -- Projektnummer 279384907 -- 
SFB 1245 and by the State of Hesse within the Research Cluster ELEMENTS (Project No. 500/10.006).


%
\appendix
\section{Comment on color neutrality}
\label{app:cneutral}
In Subsec.~\ref{subsec:effact}, we introduce the effective potential~$U$ of our model which turns out to be independent of
the color chemical potential~$\mu_3$, see Eq.~\eqref{eq:effactmodel}. In this appendix, we show that this choice for~$\mu_3$ is correct for 
studies of color-neutral matter. To this end, we consider the effective potential~$U$ for finite~$\mu_3$:
{\allowdisplaybreaks
\be
 U &=&\frac{1}{2}\bar{\lambda}_{\text{csc}}^{-1}  |\bar{\Delta}|^2 + 
 \bar{\lambda}_{\text{eff}}  |\bar{\Delta}|^4 - \frac{\mu_\text{u,b}^4}{12\pi^2}- \frac{\mu_\text{d,b}^4}{12\pi^2}- \frac{\mue^4}{12\pi^2}
\nn\\ 
&& 
\quad -  4 \bar{l}(|\bar{\Delta}|^2)\Big|_{\bar{\mu}=\mu_\text{rg}}-  4 \bar{l}(|\bar{\Delta}|^2)\Big|_{\bar{\mu}=\mu_\text{gr}}\nn\\
&& 
\quad\quad + \frac{1}{2}\theta\big(\delta\mu^2_\text{rg} \!-\! |\bar{\Delta}|^2\big) \delta \bar{l}(|\bar{\Delta}|^2)\Big|_{\substack{\bar{\mu}=\mu_\text{rg} \\ \delta\mu=\delta\mu_\text{rg}}}
\nn\\
&&
\quad\quad\quad +   \frac{1}{2}\theta\big(\delta\mu^2_\text{gr} \!-\! |\bar{\Delta}|^2\big) \delta \bar{l}(|\bar{\Delta}|^2)\Big|_{\substack{\bar{\mu}=\mu_\text{gr} \\ \delta\mu=\delta\mu_\text{gr}}}.
\label{eq:effactmu3}
\ee
Here,}
\be
\delta\mu_\text{rg} = \delta\mu+\frac{1}{2}\mu_3\,,\quad
\delta\mu_\text{gr}&=&\delta\mu-\frac{1}{2}\mu_3\,,
\ee
and
\be
\mu_\text{rg}&=&\frac{\muu+\mud}{2}+\frac{\mur+\mug}{2}\nn\\
&=& \frac{\muu+\mud}{2}+\frac{1}{2\sqrt{3}}\mu_8
\label{eq:appmurg}
\ee
are chemical potentials associated with red and green quarks. Note that we have
\be
\mu_\text{gr}&=& \mu_\text{rg}
\ee
and
\be
\mu_3 = \mur-\mug\,.
\ee
The chemical potentials related to blue quarks are
\be
\mu_\text{u,b}=\muu+\mub =\muu - \frac{1}{\sqrt{3}}\mu_8\,
\ee
and
\be
\mu_\text{d,b}=\mud+\mub =\mud - \frac{1}{\sqrt{3}}\mu_8\,.
\ee
Note that, in contrast to the blue quarks, the chemical potentials of the green and red quarks do not appear explicitly in our equations as we have 
chosen to replace them with suitable combinations of~$\mu_{\text{u}}$,~$\mu_{\text{d}}$, $\mu_3$, $\mu_8$, and~$\delta\mu$, see, e.g., Eq.~\eqref{eq:appmurg}. 
Such a replacement is convenient since the red and green quarks ``mix" in the color-superconducting phase anyhow. 
Finally, we add that the functions~$\bar{l}$ and~$\delta \bar{l}$ associated with loop integrals are given in Eqs.~\eqref{eq:defl} 
and~\eqref{eq:defdl}, respectively.

Strong-interaction matter is said to be color-neutral (not to be confused with color confinement), 
if the densities of the three color charges are identical:~$n_\text{r}=n_\text{g}=n_\text{b}$.
As discussed in Subsec.~\ref{sec:gcqcd}, this can be achieved by requiring that the densities~$n_3$ and~$n_8$ (associated with the 
color chemical potentials~$\mu_3$ and~$\mu_8$) vanish:
{\allowdisplaybreaks
\be
n_8=\frac{\partial P}{\partial\mu_8}&=&\frac{1}{2\sqrt{3}}\frac{\partial P}{\partial\mur}+\frac{1}{2\sqrt{3}}\frac{\partial P}{\partial\mug}-\frac{1}{\sqrt{3}}\frac{\partial P}{\partial\mub}\nn\\
&=&\frac{1}{2\sqrt{3}}\Big(n_\text{r}+n_\text{g}-2n_\text{b}\Big)\stackrel{!}{=}0
\ee
and}
{\allowdisplaybreaks
\be
n_3=\frac{\partial P}{\partial\mu_3}&=& \frac{1}{2}\frac{\partial P}{\partial\mur}-\frac{1}{2}\frac{\partial P}{\partial\mug}\nn\\
&=&\frac{1}{2}\Big(n_\text{r}-n_\text{g}\Big)\stackrel{!}{=}0\,.
\label{eq:n3app}
\ee
Since} the effective potential~\eqref{eq:effactmu3} is invariant under \mbox{$\text{r}\leftrightarrow \text{g}$}, we have~$P(\mur,\mug)=P(\mug,\mur)$. 
Thus, the constraint~\eqref{eq:n3app} is fulfilled, if~$\mur=\mug$, which is equivalent to~$\mu_3=0$. For the purposes of the present work, 
it is therefore justified to set~$\mu_3=0$ in the effective potential~$U$, as done in Eq.~\eqref{eq:effactmodel}.

\section{Comment on strange-quark effects}
\label{app:2plus1}
Considering the case of two massless quark flavors plus a strange quark with mass~$m_{\rm s}$, 
it has been shown in Ref.~\cite{Alford:2002kj} (see also Ref.~\cite{Alford:2007xm} for a review) that, to order $m_{\rm s}^4$ 
and second order in the gap, the pressure for color-flavor locked (CFL) quark matter can 
be written as follows:
\be
P_{\text{CFL}} = P_{\text{CFL}}^{(0)}(\mu,m_{\rm s}) + c_{\Delta}  (\bar{\Delta}_{\text{gap}}^{\text{CFL}})^2 \mu^2 + \dots\,,
\label{eq:PCFL}
\ee
where $\bar{\Delta}_{\text{gap}}^{\text{CFL}}$ is the CFL gap, $P_{\text{CFL}}^{(0)}$ is the pressure of unpaired quark matter including terms depending explicitly on the strange quark mass, 
and~$c_{\Delta}$ is a positive constant. 
Note that we tacitly assume that the gap divided by the chemical potential is sufficiently large, measured relative to $(m_{\rm s}/\mu)^2$, 
such that the CFL phase is favored over unpaired quark matter~\cite{Alford:2002kj}. Moreover, as done in Ref.~\cite{Alford:2002kj}, we have dropped the 
leading-order strange-quark mass correction~$\sim m_{\rm s}^2/\mu^2$ to the CFL gap.

At least at high densities, where the leading-order expression~\eqref{eq:PCFL} may be expected to be valid, it follows that  
the dependence of the pressure on the gap in the CFL phase is qualitatively the same as in the case of 
strong-interaction matter with only two massless quark flavors, see Eq.~\eqref{eq:Pwclimit}. 
Assuming now that the CFL gap exhibits the typical BCS-type dependence on the chemical potential and that~$\bar{\Delta}_{\text{gap}}^{\text{CFL}}(\mu)/\mu\!\to\! 0$ 
for \mbox{$\mu\to \infty$} (as it is the case for the 2SC gap), we cautiously conclude that the speed of sound in the CFL phase also
exceeds the asymptotic value~$c_{\rm s}=1/\sqrt{3}$ and exhibits 
a density dependence similar to the one in Eq.~\eqref{eq:cssana}, at least at sufficiently high densities. Recall that the density dependence of the 
speed of sound given in Eq.~\eqref{eq:cssana} follows from the expansion~\eqref{eq:Pwclimit} for the pressure.

\bibliography{qcd}

\end{document}